\documentclass[aps,pra,10pt,twocolumn,superscriptaddress,showpacs,preprintnumbers,amsmath,amssymb]{revtex4-1}

\usepackage{color}
\usepackage{graphicx}
\usepackage{bm}
\usepackage{dcolumn}
\usepackage{verbatim}
\usepackage{mathbbol}
\usepackage{makeidx}
\usepackage{amsmath}
\usepackage{placeins}

\definecolor{dark_red}{rgb}{0.75,0,0}
\definecolor{dark_purple}{rgb}{0.75,0,0.75}
\definecolor{dark_blue}{rgb}{0,0,0.75}
\definecolor{dark_green}{rgb}{0,0.60,0}
\usepackage[colorlinks,linkcolor=dark_green,citecolor=blue,linkcolor=dark_green]{hyperref}

\begin{document}

\def \scale {0.4}
\def \scaletwo {0.15}
\newcommand{\dr}[1]{\textcolor{dark_red}{#1}}
\newcommand{\dpu}[1]{\textcolor{dark_purple}{#1}}
\newcommand{\db}[1]{\textcolor{dark_blue}{#1}}
\newcommand{\orr}[1]{\textcolor{blue}{\bf[#1]}}
\newcommand{\bpsi}{\boldsymbol{\psi}}
\newcommand{\sumint}{\sum\hspace{-12pt}\int}
\newcommand{\h}[1]{\phantom{#1}}
\newcommand{\vsep}{\vspace{0.4cm}\noindent}
\newcommand{\bws}[0]{\hspace{-2mm}}
\newcommand{\pnt}[1]{{\bf\dr{$\bullet$\,\,#1\,\,$\bullet$\,\,}}}

\newcommand{\SCF}{{\scriptscriptstyle\mathrm{SCF}}}
\newcommand{\SCFp}{{\scriptscriptstyle\mathrm{SCF}^+}}
\newcommand{\XUV}{{\scriptscriptstyle\mathrm{XUV}}}
\newcommand{\IR}{{\scriptscriptstyle\mathrm{IR}}}
\newcommand{\RABITT}{{\footnotesize\textsc{RABITT }}}
\newcommand{\APT}{{\scriptscriptstyle\mathrm{APT}}}
\newcommand{\CAP}{{\scriptscriptstyle\mathrm{CAP}}}

\newcommand{\sjs}[6]{
   \left\{
       \begin{array}{ccc}
          \bws#1 & \hspace{-3mm}#2 &\hspace{-3mm} #3 \bws\\
          \bws#4 & \hspace{-3mm}#5 &\hspace{-3mm} #6 \bws
       \end{array}
   \right\}
}

\title{Attosecond photoelectron spectroscopy of helium doubly excited states}

\newcommand{\ucf}{Department of Physics \& CREOL, University of Central Florida, Orlando, Florida 32816, USA}
\newcommand{\anuc}{
Department of Physics, Stockholm University,
  AlbaNova University Center, SE-106 91 Stockholm, Sweden, EU
}
\author{Luca Argenti}\email{luca.argenti@ucf.edu}
\affiliation{\ucf}
\author{Eva Lindroth}
\affiliation{\anuc}

\date{\today}

\begin{abstract}
We describe a numerical method that simulates the interaction of the helium atom with sequences of femtosecond and attosecond light pulses. The method, which is based on the close-coupling expansion of the electronic configuration space in a B-spline bipolar spherical harmonic basis, can accurately reproduce the excitation and single ionization of the atom, within the electrostatic approximation. The time dependent Schr\"odinger equation is integrated with a sequence of second-order split-exponential unitary propagators. The asymptotic channel-, energy- and angularly-resolved photoelectron distributions are computed by projecting the wavepacket at the end of the simulation on the multichannel scattering states of the atom, which are separately computed within the same close-coupling basis. This method is applied to simulate the pump-probe ionization of helium in the vicinity of the $2s/2p$ excitation threshold of the He$^+$ ion. This work confirms the qualitative conclusions of one of our earliest publications [L Argenti and E Lindroth, Phys. Rev. Lett. {\bf 105}, 53002 (2010)], in which we demonstrated the control of the $2s/2p$ ionization branching-ratio. Here, we take those calculations to convergence and show how correlation brings the periodic modulation of the branching ratios in almost phase opposition. The residual total ionization probability to the $2s+2p$ channels is dominated by the beating between the $sp_{2,3}^+$ and the $sp_{2,4}^+$ doubly excited states, which is consistent with the modulation of the complementary signal in the $1s$ channel, measured in 2010 by Chang and co-workers~[S Gilbertson~\emph{et al.}, Phys. Rev. Lett. {\bf 105}, 263003 (2010)].
\end{abstract}

\pacs{32.80.Fb,\,\,32.80.Rm,\,\,32.80.Zb}

\maketitle

\section{Introduction}

A critical aspect of electronic motion in matter is its correlated character~\cite{Hattig2011,Sansone2012}. By avoiding each other, electrons reduce their mutual repulsion. Such correlation energy is comparable to the energy variation in many chemical processes at equilibrium. Photoelectron spectroscopies have been particularly useful to investigate correlated electronic states and their dynamics in atoms and molecules. Third-generation synchrotron, for example, generate highly monochromatic extreme ultraviolet (XUV) and x-ray radiation that enables the measurement of ionization spectra in stationary regime and with high energy resolution~\cite{Bilderback2005,Wuilleumier2006,DeOliveira2011,Patanen2015,Schippers2015,JaeschkeReference2016,Piancastelli2017,YabashiTanaka2017}. These measurements, however, provide only limited information on how the photoemission process unfolds in time.

The development of sources of sub-femtosecond XUV light pulses~\cite{Sansone2006,Goulielmakis2008} has opened the way to study photoemission from atoms, molecules and solids in a time resolved way~\cite{Lepine2013,Lepine2014,Leone2014,LeoneNeumark2016,Nisoli2017}. Today, the generation of attosecond light pulses~\cite{Chini2014} is realized in several laboratories worldwide and it allows experimentalists to implement pump-probe excitation schemes that can access electron dynamics in atoms and molecules at its natural time scale~\cite{Krausz2009,Popmintchev2012}. These schemes are used in association with either photo-fragment coincidence-detection techniques, such as COLTRIMS~\cite{Ullrich2003a,Fuchs2019}, VMI~\cite{Eppink1997}, and magnetic bottles~\cite{Kotur2016,Gruson2016}, or with high-resolution spectrometers for the ionizing light that is trasmitted through the sample~\cite{OttNature2014}. For example, attosecond pump-probe schemes can detect the minuscule relative delay with which electrons are emitted from different shells~\cite{Schultze2010a,Dahlstrom2013}, or in different directions~\cite{Heuser2016,Cirelli2018,Fuchs2019}. These delays are associated to the fast transit of the photoelectron out of the atom, and to the dynamical response of the other electrons in the residual parent ion~\cite{PazourekRMP2015,Ossiander2016a}.  Furthermore, time-resolved experiments are essential to devise quantum-control protocols of electronic excitations~\cite{Johnsson2007}, which also take place on a sub-femtosecond time scale. The ultra-short pulses and strong probe fields used in attosecond measurements activate high-order non-stationary ionization regimes. The theoretical interpretation of such experiments, therefore, often requires to solve the time-dependent Schr\"odinger equation (TDSE) numerically~\cite{Hu2006,Morishita2007,Palacios2009,Argenti2010}.  Due to the highly correlated character of multiply excited electronic states, the theoretical models used to represent poly-electronic systems often need to go beyond the single-active-electron approximation. 

The helium atom is an ideal benchmark to study the effect of electronic correlation~\cite{Argenti2010,OttNature2014,Argenti2014,Mehmood2021}. In this work we describe some of the capabilities of a set of programs for the time-resolved description of the ionization of helium through the interaction with a sequence of light pulses. The program defines a numerical close-coupling space for the single-ionization sector of the Hilbert space of the atom in a arbitrary symmetry. It builds bound as well as multi-channel scattering states of the atom within the electrostatic approximation. It computes the evolution of an arbitrary initial state under the influence of external pulses by solving the TDSE numerically, and it extracts the asymptotic observables by carrying out a spectral analysis of the resulting wavepacket. Attosecond pump-probe photoelectron spectroscopies do not directly access the time evolution of a localized wavefunction. Instead, they detect the photo-fragments after their mutual interaction and their interaction with the radiation field is over. Projecting the wavepacket obtained from a TDSE simulation on scattering states that fulfill incoming boundary conditions~\cite{Madsen2007,Argenti2010,Argenti2013}, therefore, is an efficient way to determine the asymptotic photo-fragment distribution, resolved by channel, energy and photoemission angle. This approach is particularly suited when autoionizing states are involved~\cite{Lindroth2012a}, since it does not require to wait for their complete decay to the continuum. 

The TDSE solver, which is implemented in a parallel version based on the PETSc MPI-interface library~\cite{petsc-web-page,petsc-user-ref,petsc-efficient}, has been used to benchmark the soft-photon approximation to describe atomic ionisation with trains of attosecond pulses in association with moderately strong IR pulses~\cite{Jimenez2013}, the role of intermediate autoionising states in RABITT spectroscopy~\cite{Jimenez2014,Jimenez2016}, the attosecond transient absorption spectrocopy of doubly excited states~\cite{Argenti2015b,Petersson2017}, to guide the reconstruction and control of coherent metastable doubly-excited wave packets in helium~\cite{OttNature2014}, and to control the coherence of the residual He$^+$ parent ions~\cite{Mehmood2021}. The method employed here to extract the asymptotic photoelectron distribution from time-dependent correlated wave packets has been successfully transferred to the analysis of helium wave functions expressed in basis other than B-splines~\cite{Argenti2013} and it has been extended to more complex poly-electronic atoms as well~\cite{Carette2013}. In our past work~\cite{Argenti2010}, we demonstrated that coherent excitation of doubly excited states in helium could be used to control the branching ratio between the $2s$ and $2p$ shake-up channels, $P_{2s}/P_{2p}$. In this work we carry out the demanding calculations needed to bring the simulations to full convergence. The new simulations not only confirm the qualitative conclusions in~\cite{Argenti2010}, but also show how the excursion of the branching ration is even more pronounced when correlation is fully taken into account. Furthermore, the new results for the ionization of helium with excitation of the He$^+$ parent ion to the $n=2$ states exhibit modulations that compare favorably with the measurements of the almost complementary $1s$ ionization probability near the dominant $sp_{2,2}^+$, by the group of Z. Chang~\cite{Gilbertson2010}. This finding suggests that Chang and co-workers may have detected the coherent excitation of multiple doubly excited states in helium years before the optical measurements by Ott~\emph{et al.}~\cite{OttNature2014}.

This paper is organized as follows. Section~\ref{sec:theory} describes the theoretical methods used to define the close-coupling basis, the simulation, and the wave-packet analysis. Section~\ref{sec:results} illustrates the capability of the method by examining in detail the fully resolved photoelectron distribution of the ionization of helium by a single XUV attosecond pump pulse in association with a moderately strong IR probe pulse. Section~\ref{sec:conclusions} summarizes the conclusions of this work.

\section{theory}\label{sec:theory}
The states of the helium atom are expanded in a basis that comprises a set of close-coupling (CC) channel functions~(see \cite{Argenti2006,Argenti2013} and references therein), as well as a set of confined two-electron configurations (localized channel, or LC). The CC set accounts for the long-range part of both single-ionization states and of highly-excited Rydberg satellites, whereas the LC completes the description of a state's correlated character within a short distance from the nucleus, typically of the order of ten Bohr radii. Such representation is appropriate for dynamical regimes in which only single-ionization processes take place at an appreciable rate. 
In each close-coupling channel, a bound hydrogenic state $a$ of the He$^+$ parent ion, with angular momentum $L_a$ and principal quantum number $N_a$, is coupled to a single-electron function, with well defined orbital angular momentum $\ell$ and arbitrary reduced radial component $f_i(r)$,  to give rise to a state with well defined total multiplicity $2S+1$, spin projection $\Sigma$, parity $\Pi$, angular momentum $L$, and angular momentum projection $M$, collectively referred to by the symmetry label $\Gamma = (\Pi,S,\Sigma,L,M)$,
\begin{equation}\label{eq:ccstate}
\begin{split}
\phi^{\Gamma}_{\alpha i}(\mathbf{x}_1,\mathbf{x}_2) &= \Theta_{S\Sigma}(\zeta_1,\zeta_2)[1(-)^{S}\mathcal{P}_{12}]\\
&\mathcal{Y}^{LM}_{L_a\ell_\alpha}(\Omega_1,\Omega_2)R_{N_aL_a}(r_1)\frac{f_{\ell_\alpha i}(r_2)}{r_2}.
\end{split}
\end{equation}
The channel label $\alpha$ identifies the state $a$ of the parent ion and the orbital angular momentum $\ell_\alpha$ of the photoelectron. In a channel state, the orientation of the parent ion is generally undefined, due to its coupling with the photoelectron. In~\eqref{eq:ccstate}, $x_i\equiv(\vec{r}_i,\zeta_i)$ identifies the spatial and spin coordinates of the $i$-th electron, $\Omega_i$ its spherical coordinates, $\Omega_i=(\theta_i,\varphi_i)$, $\mathcal{P}_{12}$ is the permutation of the two electron coordinates, $R_{n\ell}(r)$ is the radial part of the bound hydrogenic state with principal quantum number $n$, orbital quantum number $\ell$, normalized as $\int_0^{\infty} dr\, r^2 R^2_{n\ell}(r) = 1$~\cite{MessiahQM}. The functions $\mathcal{Y}^{LM}_{L_a\ell_\alpha}(\Omega_1,\Omega_2)$ are bipolar spherical harmonics~\cite{Varshalovich},
\begin{equation}
\mathcal{Y}^{LM}_{\ell_1\ell_2}(\Omega_1,\Omega_2)=
\sum_{m_1 m_2} C_{\ell_1 m_1, \ell_2 m_2}^{LM} Y_{\ell_1 m_1}(\Omega_1) Y_{\ell_2 m_2}(\Omega_2),
\end{equation}
where $Y_{\ell m}(\Omega)$ are ordinary spherical harmonics and $C_{\ell_1 m_1,\ell_2 m_2}^{LM}$ are Clebsch-Gordan coefficients~\cite{Varshalovich}. The two-electron spin function
$\Theta_{S\Sigma}(\zeta_1,\zeta_2)$ is defined as
\begin{equation}
\Theta_{S\Sigma}(\zeta_1,\zeta_2) = \sum_{\sigma_1 \sigma_2} C_{1/2 \sigma_1, 1/2 \sigma_2}^{S\Sigma} {^2\chi_{\sigma_1}(\zeta_1)}\,{^2\chi_{\sigma_2}(\zeta_2)},
\end{equation}
or, more prosaically, $\Theta_{11}=\alpha\alpha$, $\Theta_{1-1}=\beta\beta$, $\Theta_{10}=(\alpha\beta+\beta\alpha)/\sqrt{2}$, $\Theta_{00}=(\alpha\beta-\beta\alpha)/\sqrt{2}$, with $\alpha(\zeta)={^2\chi_{1/2}(\zeta)}=\delta_{\zeta,1/2}$, and $\beta(\zeta)={^2\chi_{-1/2}(\zeta)}=\delta_{\zeta,-1/2}$. 
It is well known that a truncated set of close-coupling channel functions~\eqref{eq:ccstate} does not give rise to a complete basis for two-electron states, even when only bound or single-ionization states well below the double-ionization threshold are considered. This is because parent-ion bound states cannot reproduce the sharp spatial modulation of a poly-electronic wave function in proximity of the coalescence of two electrons. The latter behavior, however, can be recovered by including in the basis a complementary set of symmetry-adapted two-electron configurations ${^{2S+1}(n_1\ell_1,n_2\ell_2)_{LM\Sigma}}$,
\begin{equation}\label{eq:lcFunction}
\begin{split}
\varphi^\Gamma_j(\mathbf{x}_1,\mathbf{x}_2)&=N_j\,\Theta_{S\Sigma}(\zeta_1,\zeta_2)\,[1(-)^{S}\mathcal{P}_{12}]\,\times\\
&\times
\mathcal{Y}^{LM}_{\ell_1\ell_2}(\Omega_1,\Omega_2)\,\frac{\varphi_{n_1 \ell_1}(r_1)\,\varphi_{n_2 \ell_2}(r_2)}{r_1r_2},
\end{split}
\end{equation}
where the reduced radial orbitals $\varphi_{n\ell}(r)$ are confined within a radius that both electrons can reach with appreciable probability, and $N_j$ is a normalization constant.
To summarize, any bound or single-ionization two-electron wave function with total symmetry $\Gamma$, $\Psi^\Gamma(\mathbf{x}_1,\mathbf{x}_2)$ can be well approximated by the following close-coupling expansion (we omit the variables, for brevity)
\begin{equation}\label{eq:ccSymGeneric}
\Psi^\Gamma = \sum_{\alpha i} \phi^\Gamma_{\alpha i}\,\, c_{\alpha i}^\Gamma + \sum_j \varphi_j^\Gamma\,\,b_j^\Gamma,
\end{equation}
where $c_{\alpha i}^\Gamma$ and $b_j^\Gamma$ are in general complex coefficients.

To predict the configuration and energies of helium stationary states, as well as their free and light-driven evolution, we must specify the model for the field-free Hamiltonian $H_0$, as well as the time-dependent interaction Hamiltonian $H_I(t)$.
In this work we will assume the electrostatic approximation for $H_0$,
\begin{equation}\label{eq:H0}
H_0=\frac{p_1^2}{2}+\frac{p_2^2}{2}-\frac{2}{r_1}-\frac{2}{r_1}+\frac{1}{r_{12}},
\end{equation} 
since relativistic effects play only a minor role for the short-time evolution of an atom as light as helium.
Indeed, the spin-orbit splitting in helium manifests itself only on a time scale of several picoseconds, whereas the pump-probe ionization processes we are interested in take place on a time scale that is at least two orders of magnitude shorter. Within the electrostatic approximation, all the quantum numbers in $\Gamma$ (parity, total spin, total angular momentum, and their projections) are conserved.

\paragraph{B-spline basis.}
The reduced one-electron radial functions in Eq.~\eqref{eq:ccstate} and~\eqref{eq:lcFunction} are represented here as linear combinations of B-splines~\cite{deBoor}. B-splines form a flexible set of compact-support functions that has proven ideally suited to represent bound, Rydberg, and continuum orbitals in atomic and molecular physics~\cite{Bachau2001}.
B-splines are a basis for piecewise polynomials of degree $n$ that are $\mathcal{C}^\infty$ everywhere except for an assigned discrete set of nodes $\{t_i\}$, in correspondence of which the $n$-th derivative of the B-splines can be discontinuous. It is common practice to label B-splines by their order $k=n+1=1,\,2,\,\ldots$, rather than their polynomial degree $n$. The set of $k-$th order B-splines $\{B^k_i(x)\}$ can be defined recursively as~\cite{deBoor},
\begin{eqnarray}
B_i^1(x)&=&\theta(x-t_i)\,\theta(t_{i+1}-x),\label{eq:bsdef1}\\
B_i^k(x)&=&\frac{(x-t_i)\,B_i^{k-1}(x)}{t_{i+k-1}-t_i}\,+\,\frac{(t_{i+k}-x)\,B_{i+1}^{k-1}(x)}{t_{i+k}-t_{i+1}},\label{eq:bsdef2}
\end{eqnarray}
where $\theta(x)=\int_{-\infty}^x\delta(\tau)d\tau$ is the Heaviside step function.
Some intervals $[t_i,t_{i+1}]$ can have zero length, giving rise to nodes with multiplicity $\nu$ higher than one, $t_i=t_{i+1}=\cdots=t_{i+\nu-1}$, in which case the B-splines can exhibit a discontinuity at $x=t_i$ in the derivative of order as high as $(k-\nu)$. In particular, for  B-splines to assume finite values at the boundaries of the representation interval, the first and last nodes must be $k$ times degenerate. Any B-spline of order $k$ differs from zero only in an interval delimited by $k+1$ consecutive nodes, when the nodes are counted with their multiplicity. Thanks to this property, local linear operators in a B-spline basis have a banded representation, which helps making the algorithms that solve linear systems in this basis numerically stable. Here, B-splines-order $k=10$ is used, which gives a good balance between representation accuracy and numerical stability. Furthermore, the present computational scheme employs two different B-spline bases: a small one for the localized orbitals and a large one for the diffuse and continuum orbitals. The localized B-spline basis is defined in terms of a radial grid where the separation between consecutive nodes increases linearly at moderate distances, until it reaches a radius comparable to the size of the most energetic ion that can be excited during photoionization (typically of the order of several tens of Bohr radii), and which is suited to describe both the parent-ion states and the localized channel. In the small B-spline basis, consecutive nodes are approximately $6$~a.u. aparts at a distance of $\sim$40~a.u. from the nucleus. The large B-spline basis is defined by a radial grid that comprises all the points in the smaller grid, plus all those needed to give rise to a uniform asymptotic separation  between consecutive nodes (typically of the order of 0.5~a.u.) and to reach distances of the order of several hundreds or even thousands Bohr radii. With two different B-spline bases, it is possible to drastically reduce the number of configurations needed to reproduce the effects of correlation at short range. For example, in the localized region, the large B-spline basis comprises three times as many functions as the localized B-spline basis. Since the set of nodes for the localized B-spline set is chosen as a subset of the nodes that define the continuum B-spline basis, the localized B-spline space is an exact subspace of the continuum B-spline space~\cite{Argenti2006}.

The reduced radial part $\varphi_{n\ell}(r)$ of an orbitals with angular momentum $\ell$ must comply with the regularity condition at the origin $\lim_{r\to 0} \varphi_{n\ell}(r) / r^{\ell} = 0$. In the present work, this condition is explicitly satisfied for $\ell \leq 4$ by eliminating the first $\ell+1$ B-splines from the basis,
\begin{equation}
\varphi_{n\ell}(r) = \sum_{i>\ell+1} B_{i}(r) \,c_{i;n\ell}.
\end{equation}

\paragraph{Basis conditioning.}

The close-coupling states and the localized states as defined in~\eqref{eq:ccstate} and~\eqref{eq:lcFunction} give rise to redundant configurations. For example, in the construction of the $2s\varepsilon_p$ $^1$P$^o$ CC channel, the {$2s$} parent-ion state can be coupled to a {$2p$} spin-orbital state for the outer electron, to give rise to a $2s2p$ $^1$P$^o$ doubly-excited configuration. The same configuration, however, is generated also in the construction of the $2p\varepsilon_s$ $^1$P$^o$ CC channel, as well as in the localized channels, which comprises all  $ns\, mp$ configurations. To avoid these redundancies, the orbital space for the outer electron in the close-coupling channels, as well as the CI space of the localized channel, must be restricted. In the following, we summarize the conventions and procedures followed here.

The parent-ion orbitals are computed by diagonalizing the hydrogenic He$^+$ Hamiltonian in the localized B-spline basis.
In the construction of the CC channels~\eqref{eq:ccstate}, a parent ion with principal quantum number $N_a$ and orbital angular momentum $L_a$ is coupled only to those orbitals from the continuum B-spline space that are orthogonal to all the parent-ion orbitals with $n<N_a$, as well as to the parent ions with $n=N_a$ and $\ell< L_a$. In this way, the close-coupling channels are rigorously orthogonal. For example, the $1s2p$ configuration obtained by coupling a $1s$ ion to a $2p$ He$^+$ hydrogenic orbital is represented in the $1s\epsilon_p$ channel, but not in the $2p\epsilon_s$ channel.

The localized-channel configurations, which are meant to reproduce short-range correlation, are built from the self-consistent-field (SCF) orbitals that diagonalize an effective one-particle Hamiltonian given by the Hartree-Fock operator of the $1s^2$ ground state of helium plus an additional single-charge Coulomb attraction potential that operates on the space orthogonal to the $1s$ orbital. It should be noted that, in the present work, the spectral analysis of the Hamiltonian and the time propagation are conducted in a close-coupling space that comprises the full configuration-interaction (full-CI) basis built from the localized SCF orbitals. The full-CI basis is invariant under any unitary transformation of the localized orbital space, and hence the use of a separate set of orbitals for the ions and the LC space becomes immaterial. Still, treating hydrogenic and SCF orbitals separately offers the latitude to truncate the orbital basis in a physically meaningful way, if needed.
To avoid redundancies between the CC and the LC channels, the projector $\hat{P}$ on the CC channels,
\begin{equation}
P= \sum_{\alpha} P_\alpha,\quad\mathrm{where}\quad
P_{\alpha}= \sum_{i} |\phi_{\alpha i}\rangle\,\langle\phi_{\alpha i}|,
\end{equation}
is diagonalized on the LC basis. The states whose eigenvalue differs from $1$ less than a prescribed small threshold $\epsilon$ are eliminated from the LC space, since they are already accurately represented by the CC basis by definition. The resulting conditioned LC space (CLC) and the CC space are numerically linearly independent. When the LC is the full-CI localized space, the LC is the direct sum of a CC-space subset, spanned by those configurations with at least one parent-ion orbital represented in the CC series, and of a space orthogonal to all CC functions. The CLC and CC space, therefore, are not merely independent, they are orthogonal.

\paragraph{Box eigenstates}
Let us indicate with $|\boldsymbol{\varphi}^\Gamma\rangle=(\varphi^\Gamma_1,\varphi^\Gamma_2,\ldots)$ the row vector of the localized configurations, with $|\boldsymbol\phi_\alpha^\Gamma\rangle=(\phi_{\alpha,1},\phi_{\alpha,2},\ldots,\phi_{\alpha,N^\Gamma_\alpha})$ the row vector of states with symmetry $\Gamma$ and in the partial wave-channel $\alpha$, and with $|\boldsymbol\bar{\phi}_\alpha^\Gamma\rangle = (\bar{\phi}_{\alpha,1},\bar{\phi}_{\alpha,2},\ldots,\bar{\phi}_{\alpha,N_\alpha^\Gamma-1})$ the vector of channel states obtained by eliminating from the radial basis the last B-spline, i.e., the only B-spline that does not vanish at the box boundary.
To compute the bound states with symmetry $\Gamma$, we diagonalize the Hamiltonian in the basis $|\boldsymbol{\phi}^{\Gamma}_{{\rm b}}\rangle=|\boldsymbol{\varphi}^\Gamma\rangle \oplus \bigoplus_{\alpha} |\boldsymbol{\bar{\phi}}_\alpha^\Gamma\rangle$, which vanishes at the box boundary (and hence, the matrix representation of the Hamiltonian is Hermitian),
\begin{equation}
H_0|\boldsymbol{\psi}^\Gamma_{\rm b}\rangle = |\boldsymbol{\psi}^\Gamma_{\rm b}\rangle \, \boldsymbol{E}^\Gamma_{\rm b},\quad 
|\boldsymbol{\psi}^\Gamma_{\rm b}\rangle = |\boldsymbol{\phi}^{\Gamma}_{{\rm b}}\rangle\,\mathbf{c}^\Gamma_{\rm b}.
\end{equation}
As discussed in Sec.~\ref{sec:TDSE}, the discrete spectral representation $\{(|\boldsymbol{\psi}^\Gamma_{\rm b}\rangle,\,\boldsymbol{E}^\Gamma_{\rm b})\}_\Gamma$ of the field-free Hamiltonian is used both to define the bound states of the atom and to evaluate the time propagator of the system driven by external fields.

\paragraph{Scattering states.}

Many equivalent methods exist to solve the multi-channel secular problem for helium and more complex atoms represented in a B-spline basis, within the close-coupling \emph{ansatz}~\cite{Nikolopoulos2001,Zatsarinny2013a,Carette2013,Argenti2013,Marante2017}.
In the present work, the single-ionization scattering states of helium below the double ionization threshold are computed using the B-spline K-matrix method~\cite{Cacelli1986,Cacelli1991}. The K-matrix method is an $L^2$ realization of a configuration-interaction calculation in the continuum that has been successfully applied to a number of problems in atomic and molecular physics~\cite{Moccia1991,Mengali1996,Fang2000,Argenti2006,Argenti2007,Argenti2007a,Argenti2008b,Argenti2010b,Lindroth2012a,Argenti2016}. This method will be briefly summarized below in the case of helium, for scattering states with total symmetry $\Gamma$, which we will not explicitly indicate, for brevity.

First, the total Hamiltonian $H_0$ is diagonalized in the subspace of each close-coupling channel $|\boldsymbol{\bar{\phi}}_{\alpha}\rangle$ confined to the quantization box, giving rise to so-called partial-wave channel (PWC) states $|\phi_{\alpha \epsilon_i}\rangle$,
\begin{equation}\label{eq:pwc}
\langle \phi_{\alpha \epsilon_i} | H_0 | \phi_{\alpha \epsilon_j} \rangle = \delta_{ij}\, \epsilon_i \,\langle \phi_{\alpha \epsilon_i} | \phi_{\alpha \epsilon_i} \rangle.
\end{equation}
The eigenvalues $\epsilon_i$ below the channel threshold $E_a$ (i.e., the parent-ion energy) approximate the first few terms of a discrete Rydberg series that converges to the threshold. They correspond to an electron with fixed orbital angular momentum bound by the He$^+$ ion frozen in a given state $a$. The eigenvalues above threshold, $\epsilon_i > E_a$ are a discrete selection of energies from the continuum spectrum, $E>E_a$. They correspond to those stationary states in which an asymptotically free electron scatters elastically off the parent ion, without changing angular momentum, and whose incoming and outgoing component happen to interfere destructively at the box boundary. The positive-energy PWC functions exhibit a channel phase shift $\delta_{\alpha i}$ with respect to the regular Coulomb function with the same orbital angular momentum and asymptotic energy. These discretized-continuum states are rescaled so that their analytic extension at arbitrarily large radii fulfills the normalization condition $\langle \phi_{\alpha E} | \phi_{\alpha E'} \rangle = \delta(E-E')$. In this way, we can define an elastic-scattering Hamiltonian $H_0$ as 
\begin{equation}\label{eq:srH0}
H_0 = \sum_{\gamma} \sumint d\epsilon |\phi_{\gamma \epsilon}\rangle \epsilon \langle \phi_{\gamma \epsilon} |,
\end{equation}
where the index $\gamma$ runs over both open and closed PWC's, as well as on the CLC (which comprises only discrete states). 
Under the assumption that the elastic channels are asymptotically decoupled, the elastic-scattering Hamiltonian $H_0$ acts on single-ionization wave packets at large distance from the nucleus in exactly the same way as the full Hamiltonian $H$. Conversely, the perturbation $V=H-H_0$ acts at short range, and hence it is possible to seek the multichannel scattering eigenstates of $H$ as solutions of the Lippmann-Schwinger equation (LSE), e.g., in its principal-part formulation~\cite{Newton}, 
\begin{equation}
\psi_{\alpha E}^\mathcal{P} = \phi_{\alpha E} + G_0^{\mathcal{P}}(E) V \psi_{\alpha E}^\mathcal{P},\quad G_0^\mathcal{P}(E) \equiv \frac{\mathcal{P}}{E-H_0},
\end{equation}
where the index $\alpha$ runs over the channels which are open at the energy $E$.
By inserting the spectral resolution~\eqref{eq:srH0} of $H_0$, the LSE becomes
\begin{equation}
\psi_{\alpha E}^{\mathcal{P}}=\phi_{\alpha E}+\sum_\gamma\sumint d\epsilon\, 
\phi_{\gamma \epsilon}\frac{\mathcal{P}}{E-\epsilon}
\mathbf{K}_{\gamma\epsilon,\alpha E},
\label{eq:tm1}
\end{equation}
where $\mathbf{K}_{\gamma\epsilon,\alpha E}=\langle \phi_{\gamma \epsilon}|V|\psi^{\mathcal{P}}_{\alpha E}\rangle$ are elements of the \emph{off-shell} reaction matrix, which is known to be a smooth function of the left continuous index $\epsilon$~\cite{Newton}. Equation (\ref{eq:tm1}) may be solved for the unknown coefficients $\mathbf{K}$ by requiring $\psi_{\alpha E}^\mathcal{P}$ to be an eigenfunction of the complete projected hamiltonian with eigenvalue $E$,
\begin{equation}
\langle\phi_{\beta E'}|\,E-\mathcal{H}\,|
\psi_{\alpha E}^{\mathcal{P}}\rangle=0\quad\forall\,\beta,E'.\label{eq:phiHpsi}
\end{equation}
This secular problem can be written as a systems of integral equations for $\mathbf{K}$,
\begin{equation}\label{eq:kmie}
K_{\beta E',\alpha E} - \sum_{\gamma\neq \beta} \sumint V_{\beta E',\gamma \epsilon}\frac{P}{\epsilon-E}K_{\gamma\epsilon,\alpha E} = V_{\beta E',\alpha E}.
\end{equation}
By interpolating the continuum-continuum matrix elements $V_{\beta E',\gamma \epsilon}$ from their values $V_{\beta \epsilon_i,\gamma \epsilon_j}$ between discretized PWC continua~\eqref{eq:pwc}, and the $K_{\gamma\epsilon,\alpha E}$ coefficients from their discretized counterpart $K_{\gamma\epsilon_i,\alpha E}$, Eq.~\eqref{eq:kmie} is converted to an algebraic set of linear equations and solved with standard linear-algebra routines. The scattering states with defined spherical symmetry and incoming boundary conditions $\psi_{\alpha E}^-$ are obtained as
\begin{equation}
\psi_{\alpha E}^-=\sum_\beta\psi_{\beta E}^{\mathcal{P}}\,\left[\frac{1}{\mathbf{1}-
    i\pi\mathbf{K}(E)}\right]_{\beta\alpha}e^{- i(\sigma_{\ell_\alpha}+\delta_\alpha-\ell_\alpha\pi/2)},
\end{equation}
where $\mathbf{K}_{\alpha,\beta}(E)\equiv\mathbf{K}_{\alpha E, \beta E}$ is the on-shell reactance matrix (\S $7.2.3$ in \cite{Newton}),  $\sigma_{\ell_\alpha}=\arg\Gamma(\ell+1-i/k)$ is the Coulomb phaseshift, and $\delta_\alpha$ is the channel phaseshift. Finally, the scattering states that correspond to Coulomb plane waves associated to a parent ion in a well defined state, are given by
\begin{equation}
\psi_{a;E\hat{\Omega}\sigma}^-=\sum_{\Gamma\ell m}
C_{L_a M_a,\ell
  m}^{LM}C_{S_a\Sigma_a,\frac{1}{2}\sigma}^{S\Sigma}
Y_{\ell m}^*(\hat{\Omega})\,\,\psi_{\alpha\ell E}^{\Gamma (-)}
\end{equation}
and possess the following normalization
\begin{equation}
\begin{split}
\langle \psi_{a;E\hat{\Omega}\sigma}^- |
\psi_{b;E'\hat{\Omega'}\sigma'}^-\rangle&=\delta_{ab}\delta_{\sigma\sigma'}\delta(E-E')\,\times\\
&\times\,\delta(\cos\theta-\cos\theta')\delta(\phi-\phi').
\end{split}
\end{equation}

\paragraph{Time-dependent propagator}\label{sec:TDSE}
The time-dependent Schr\"odinger equation for the wavefunction $\Psi(t)$ reads
\begin{equation}
i\partial_t\Psi(t) = \mathcal{H}(t) \Psi(t),
\end{equation}
where the total time-dependent Hamiltonian $\mathcal{H}(t)$ comprises a time-independent field-free component $H$ and a time-dependent interaction $H_I(t)$, 
\begin{equation}
\mathcal{H}(t) = H + H_I(t).
\end{equation}
In this work, we will assume $H_I(t)$ to be the interaction term of the minimal-coupling Hamiltonian in dipole approximation, in either velocity gauge, $H_{I}(t) = \alpha \vec{A}(t)\cdot \vec{P}$, or length gauge, $H_{I}(t)=\vec{E}(t)\cdot\vec{R}$, where $\alpha = e^2/\hbar c\approx 1/137.036$ is the fine-structure constant~\cite{Mohr2016}, $\vec{A}(t)$ and $\vec{E}(t)$ are the transverse vector potential and electric field of the external radiation impinging on the atom, $\vec{P}=\vec{p}_1+\vec{p}_2$ is the total electronic canonical momentum, and $\vec{R}=\vec{r}_1+\vec{r}_2$ is (minus) the electric dipole moment. 
The TDSE is integrated from an initial time $t_0$ to a final time $t=t_n$ in a sequence of $n$ time-steps as
\begin{equation}
\Psi(t) = \prod_{i=1,n}U_{\CAP}(t_{i}-t_{i-1})\,U(t_{i},t_{i-1}) \Psi(t_0),
\end{equation}
where $U(t+dt,t)$ is a second-order symmetrically split exponential unitary propagator,
\begin{equation}~\label{eq:unitaryPropagator}
U(t+dt,t)=e^{-i\,H\,dt/2}e^{-i\,H_I(t+dt/2) \,dt}e^{-i\,H \,dt/2},
\end{equation}
and $U_{\CAP}(dt)$ is an exponential complex-absorption evolution operator,
\begin{equation}~\label{eq:ucap}
U_{\CAP}(dt)=e^{-i\,dt H_{\CAP}}.
\end{equation}

Computing the action of the time-step evolution operator~\eqref{eq:unitaryPropagator} in the spectral basis of the confined field-free atom is quite elementary. The first and last step of the propagation amount to multiplying each expansion coefficient $c_i$ of the wave function by the corresponding phase factor $\exp(-i\,dt\,E_i/2)$. This part of the propagation is virtually instantaneous compared with the action of the driven component of the split propagator, and it takes care of the broadest spectral span of the Hamiltonian, since it accounts exactly for the whole kinetic energy, and field-free Coulomb interaction. The intermediate step of the propagation is accomplished by means of a Krylov expansion of the exponential $e^{-i\,H_I(t+dt/2) \,dt}$~\cite{Argenti2010}, which normally converges for a Krylov size of the order of twenty at most, for time steps of the order of 0.01~a.u.

The velocity gauge is numerically more convenient than the length gauge, when short-range correlation is not an issue, since the results converge more rapidly with respect to the maximum orbital angular momentum~\cite{Cormier1995}. In the case of helium, the velocity gauge offers additional advantages since parent ions with the same principal quantum number are degenerate, and hence their dipolar coupling vanishes, $\langle n \ell \| p_{1} \| n \ell' \rangle = i \langle n \ell \| [H_0,r_1] \| n \ell' \rangle = i (E_{n\ell}-E_{n\ell'}) \langle n \ell \| r_1 \| n \ell' \rangle = 0$. This circumstance eliminates the radiative coupling between close-coupling channels due to the anomalous Stark polarization of the parent ion, thus improving numerical accuracy.
The transverse component of the external electric field in the length gauge is computed from the vector potential as
\begin{equation}
\vec{E}(t)=-\alpha \partial_t \vec{A}(t),
\end{equation}
which ensures that the time integral of the electric field vanishes~\cite{Madsen2002}.

The absorption potentials are defined as
\begin{equation}
H_{\CAP}(r)= \sum_\alpha c_\alpha P_\alpha (r-R_\alpha)^{n_\alpha}\theta(r-R_\alpha).
\end{equation}
In principle, the absorption coefficients $c_\alpha$, exponent $n_\alpha$, and radius $R_\alpha$ have the flexibility of being optimized independently, to minimize unphysical reflections at the box boundary in each channel. In practice, however, a common set of parameters can lead to satisfactorily converged results. In this work, the complex-absorption potential range extends over a radial region as large as $100$ Bohr radii, $n_\alpha = 2$. In this way, reflections by the potential itself or by the box boundary of the fast photoelectrons in the $1s$ channel are negligible. 
The evaluation of the extinction factor~\eqref{eq:ucap} is significantly less demanding than the driven unitary propagation, since the absorption potential is zero in all but the last slice of the radial interval. This means that most of the eigenvalues of the absorption potential are zero.
Let
\begin{equation}
H_{\CAP}|\psi_{\mathrm{cap},i}\rangle = |\psi^{\mathrm{cap},i}\rangle \lambda_i,
\end{equation}
with $\lambda_i\neq0$ iff $i>N-N_{\mathrm{cap}}$, where $N$ is the total number of box states and $N-N_{\mathrm{cap}}\ll N$ is the size of the ker of the CAP in the box space. Then \eqref{eq:ucap} can be rewritten as
\begin{equation}
U_{\CAP}(dt)=\hat{1}+\sum_{i>N-N_{\mathrm{cap}}} |\psi^{\mathrm{cap}}_i\rangle (e^{-i\,dt \lambda_i}-1)\langle\psi^{\mathrm{cap}}_i|.
\end{equation}
The evaluation of the action of the extinction operator, therefore, requires a number of matrix-vector operations that is much smaller than that even a single Krylov iteration for the evaluation of the driven propagation step.

\subsection{Asymptotic observables}
The present approach offers a natural way to express the wave function $\Psi(t)$ in terms of a physical spectral basis. 
Determining the population of any bound state of the system as well as the probability with which the photoelectron is emitted with any given energy and direction, leaving behind the parent ion in any given state, therefore, is an easy and computationally inexpensive task. As soon as the external pulses are over, the evolution of any complex probability amplitude on an energy eigenstate $i$, $\mathcal{A}_i(t)=\langle i | \psi(t)\rangle$ changes by just a phase factor, $\mathcal{A}_i(t+\Delta t) = e^{-iE_i\Delta t}\,\mathcal{A}_i(t)$. This means that there is no need of protracting the propagation any longer after the end of the pulse. Furthermore, since the multichannel scattering states $|\psi_{\alpha E}^{\Gamma -}\rangle$ are known, it is possible to compute the asymptotic photoelectron distribution while the electronic wavepacket is still in the interaction region and, indeed, still transiently trapped in metastable states. The channel-resolved photoelectron amplitude in interaction-representation
\begin{equation}
\mathcal{A}^{\Gamma}_{\alpha E} = e^{iEt}\,\langle \psi_{\alpha E}^{\Gamma (-)} | \Psi(t)\rangle,\quad {E}(t')=0 \,\,\,\forall \,\,\,t'>t,
\end{equation}
is a time-invariant quantity~\cite{Argenti2013}. 
The photoelectron distribution, resolved by symmetry, channel and energy, therefore, is readily obtained as 
\begin{equation}
\frac{dP^{\Gamma}_{\alpha}(E)}{dE} = \left|\langle \psi_{\alpha E}^{\Gamma (-)} | \Psi(t)\rangle\right|^2,
\end{equation}
whereas the distribution resolved by state of the ion, energy, and photoemission angle, is given by
\begin{equation}
\frac{dP_{a}(E)}{dEd\Omega} = \sum_{\sigma M_a}\left|\langle \psi_{a; E\hat{\Omega}\sigma}^{-} | \Psi(t)\rangle\right|^2.
\end{equation}
Finally, from the partial photoelectron cross sections, it is possible to compute the partial integral photoionization yield in each channel, $P_a$, as 
\begin{equation}\label{eq:partialintegralyield}
P_a = {\sum_{\Gamma\alpha}}'\int_{E_a}^\infty dE \left|\langle \psi_{\alpha E}^{\Gamma (-)} | \Psi(t)\rangle\right|^2,
\end{equation}
where the prime in the last sum indicates that the summation is restricted to the total symmetries $\Gamma$ where the ion state $a$ is represented, and to the channels $\alpha$ in which the asymptotic state of the ion is $a$.

\section{Results~\label{sec:results}}

To illustrate the capabilities of the present approach, we examine the photoionisation of helium from the ground state to the energy region close to the $N=2$ excitation threshold of the He$^+$ parent ion, using a single XUV attosecond pulse in association with a moderately intense few-cycle IR probe pulse, 
\begin{equation}
\mathrm{He} (1s^2) + \gamma_{\XUV}\pm n\gamma_{\IR}\rightarrow \mathrm{He}^+_{1s,2s,2p} + e^-.
\end{equation}
In the latter equation, we indicated schematically the interaction with the IR laser field in term of the interchange of a definite number of IR photons. It is understood, however, that processes may occur which cannot be reduced to a truncated perturbative picture (e.g., ac-Stark shift, tunneling, above-the-threshold ionization, etc.). In this work, we assume that the laser pulses are always linearly polarized along the same direction, so that only natural symmetries can be populated.
The pump is an XUV Gaussian pulse with central frequency $\hbar\omega_\XUV=60.69$~eV (2.2308~a.u.), a duration of 385~as (full width at half maximum of the envelope of the intensity, fwhm$_\XUV$),  and a peak intensity $I_\XUV$=1~TW/cm$^2$. The probe is a few-cycle IR cos$^2$ pulse, with central frequency $\hbar\omega_\IR=1.55$~eV (0.057~a.u.), an entire duration of 10.66~fs (fwhm$_\IR\approx$3.77~fs), and peak intensity $I_\IR=$1~TW/cm$^2$.
\begin{figure}[hbtp!]
\includegraphics[width=\linewidth]{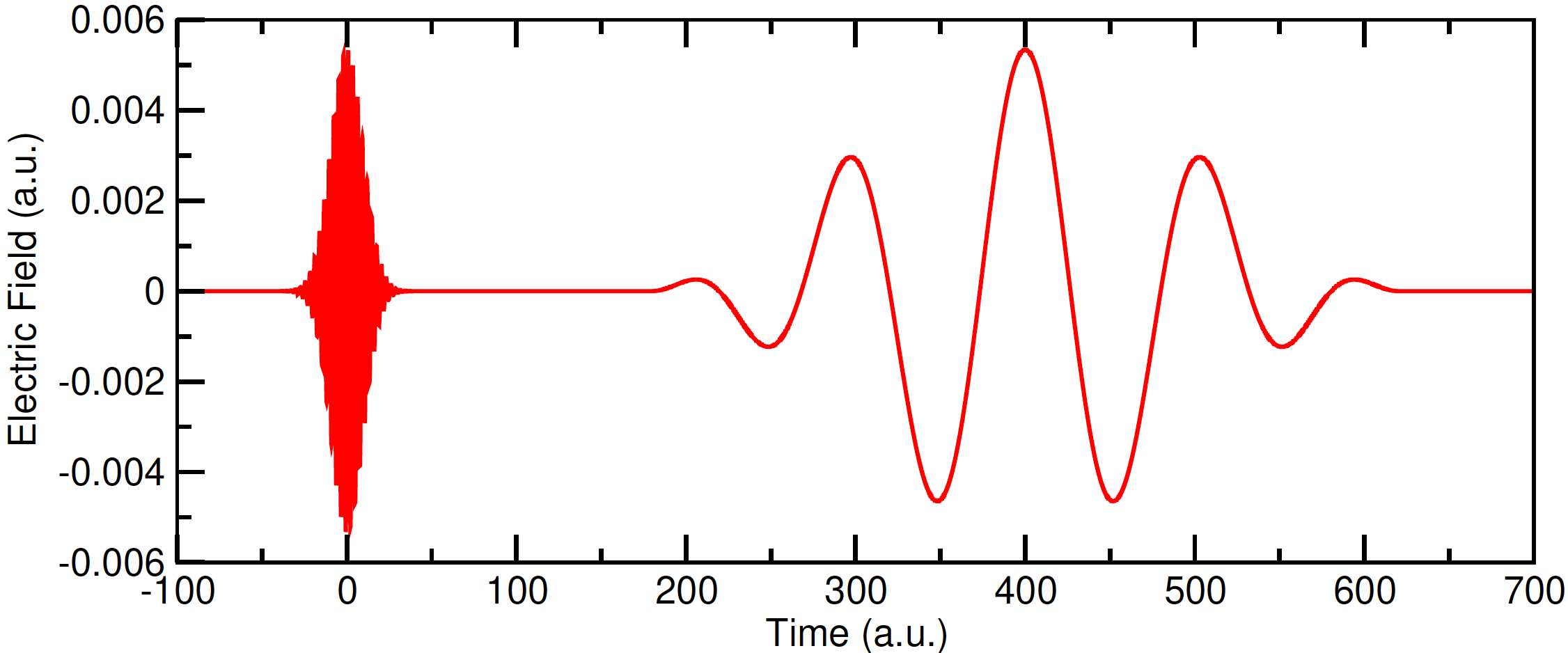}
\caption{\label{fig:PP_pulse} Example of XUV-pump-IR-probe electric field used in this work ($\tau=400$~a.u.). See text for more details.}
\end{figure}
Figure~\ref{fig:PP_pulse} illustrates the sequence of electric field pulses for a time delay $\tau=400$~a.u. (9.68~fs).

We have conducted the calculations using both a small and a larger basis set. The small basis, which comprises a minimal number of localized orbitals, reproduces the conditions used for the calculations in~\cite{Argenti2010}. Most of the electron-electron correlation beyond the essential close-coupling channels, therefore, is missing.  The larger basis comprises the full-CI pseudo-channel space, which provides most of the correlation in the energy region examined in this study.

The first set includes, for each total angular momentum $L$ up to $L=9$, the minimal-close coupling basis required to represent the $N=2$ ionization channels and the doubly-excited states converging to the $N=2$ threshold: $1s E_L$, $2s E_L$, $2p E_{L+1}$, and, for $L>0$, $2p E_{L-1}$. Beyond the close coupling part, the basis also includes the Hartree-Fock ground state $1s_\SCF^2$, which ensures that the starting point has at least HF quality, as well as the configurations $1s_\SCF 2p_{\SCFp}$, $1s_\SCF 3d_{\SCFp}$, $1s_\SCF 4f_{\SCFp}$, $1s_\SCF 5g_{\SCFp}$, $1s_\SCF 6h_{\SCFp}$, $4f_{\SCFp}^2$, which have the only purpose of ensuring that the correlation pseudo-channel is not empty. The localized parent-ion and SCF orbitals are built from the B-spline basis of order $k=10$ defined by a set of 29 non-uniformly spaced distinct radial nodes, which span from the origin to $R\simeq 41$~Bohr radii. The partial-wave radial functions are expanded on a $k=10$ B-spline basis defined by a second set of $2407$ nodes, with uniform asymptotic spacing of $0.5$~a.u. and which reaches a maximum distance of $\simeq 1200$~Bohr radii. In the representation of the single-particle wave functions with orbital angular momentum $\ell$, the first $\min(\ell+1,5)$ B-splines are eliminated to enforce the regularity of the orbitals at the origin (for $\ell>4$ this detail is not crucial). The overall size of the {$^1$L$^\pi$} spaces, with $L=0$, $1$, $2$, \dots, $9$, are: 7238, 9647, 9646, 9641, 9639, 9637, 9639, 9637, 9639, 9637. Due to the small contribution of the configurations from the partial-wave channels, the energy of the ground state is marginally better than the Hartree Fock limit: $E_{\mathrm{g}}=-2.8867\,742$~a.u. (compare with the $1s^2$ SCF energy $-2.861\,680$~a.u.).

The larger basis comprises, beyond the minimal set of close-coupling channels of the smaller basis, the full-CI set of configuration $n\ell n'\ell'$ constructed from all the localized orbitals with orbital angular momentum $\ell\leq 5$. This space provides a very good description of the short range correlation of all the states (bound, doubly excited and in the continuum) up to a total energy $E_{\mathrm{tot}}\leq -0.25$~a.u. Beyond this value, N=3 channels must be explicitly included in the close-coupling expansion to account properly for the DES converging to the N=3 threshold ($E_{N=3}=-0.222\,222$~a.u.). Notice, however, that at lower energies the optical potential reproduced by the full-CI short-range basis is perfectly capable of supplying the contribution of the missing close-coupling channels even in absence of the $N=3$ channels, and indeed of any higher single-ionization as well as double-ionization channels. The size of the short-range full-CI pseudo-channels is  1906, 3033, 3954, 3048, 2721, 1750, 1151, 552, 276, and 36, for the natural symmetries with $L=0,1,\ldots,9$, respectively.  The overall size of the {$^1$L$^\pi$} spaces, with $L=0$, $1$, $2$, $\ldots$, $9$, are: 9064, 12577, 13498, 12592, 12288, 11363, 10787, 10188, 9912, and 9672, for a total size of 111941. The energy of the ground state is $E_{\mathrm{g}}=-2.9036\,028$~a.u., to be compared with the accurate non-relativistic limit for $\ell_{\mathrm{max}}=5$, which is $-2.9036\,057$~a.u.~\cite{Carroll1979}.

\begin{figure}[hbtp!]
\includegraphics[width=\linewidth]{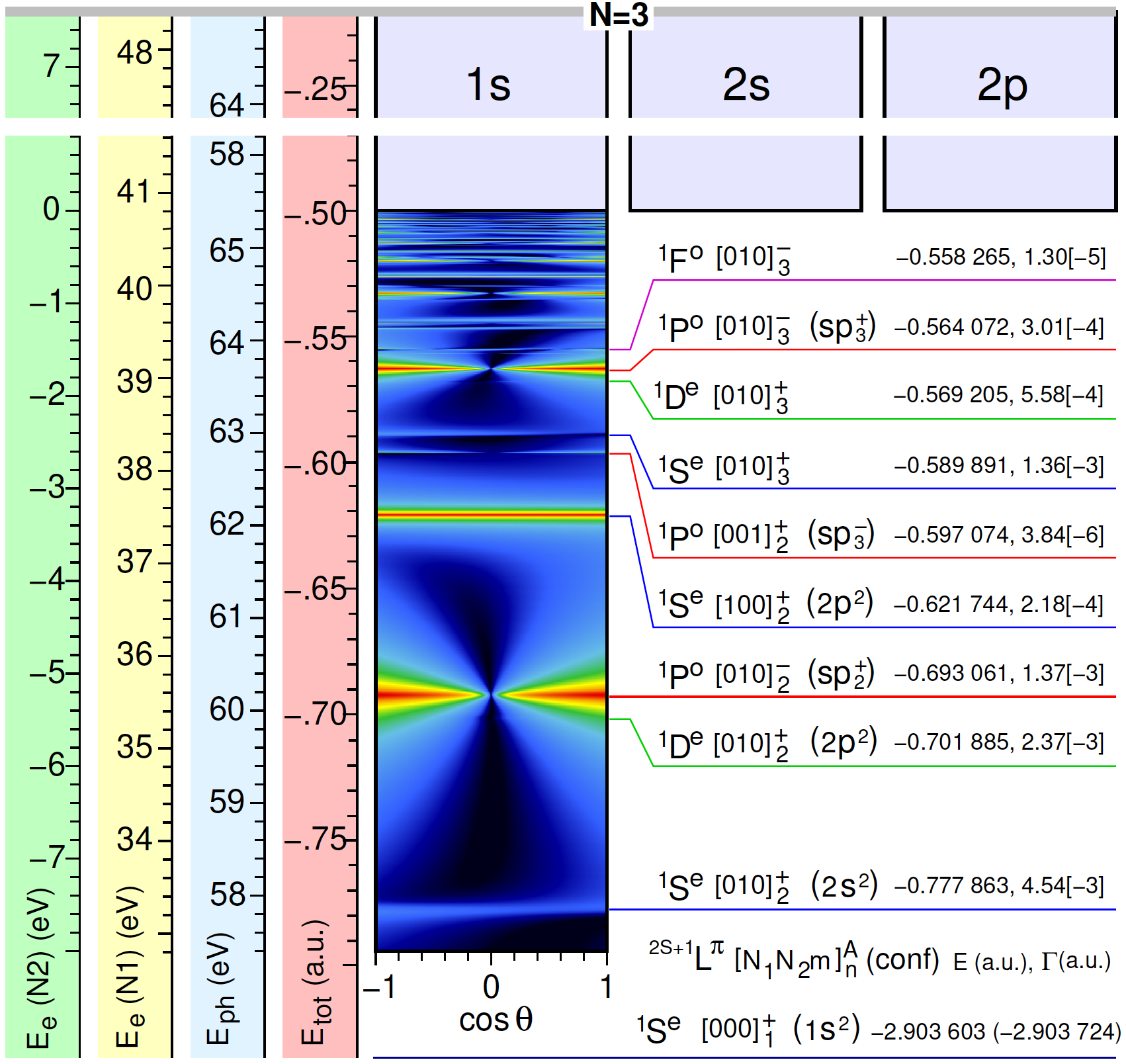}
\caption{\label{fig:EnergyScheme} Schematic energy diagram for the first few helium doubly-excited states below the $N=2$ threshold, with natural parity. The scales on the left indicate, from right to left: the total energy of the system in atomic units; the energy of the photon required to excite the state from the ground, in electron volts; the energy of the photoelectron in the $1s$ and $N=2$ channels, respectively, in electron volts. The rectangles in the upper part of the energy scheme represent the channels that $1s$, $2s$ and $2p$ that are open above the $N=2$ threshold.}
\end{figure}
Figure~\ref{fig:EnergyScheme} offers a schematic overview of the autoionizing-state levels, computed with the larger basis, in the energy region close to the N=2 threshold. Below such threshold, only the $1s$ channel is open. This means that states in the continuum will eventually lead to a $1s$ parent ion plus a free electron. The excited $2s$ and $2p$ parent-ion states can temporarily bind the second electron giving rise to doubly excited states (DES). Due to electronic repulsion, these states are coupled to the  continuum in the $1s$ channel, to which they eventually decay~\cite{Fano1961}. The DES in this energy region can be classified in terms of approximate quantum numbers. Several classification schemes have been proposed for the DESs: the original Fano's classification~\cite{Cooper1963a}, the $_N(K,T)_n^A$ scheme by Herrick and Sinanoglu, the hyperspherical scheme~\cite{Lin1974}, the Stark quantum numbers $[N_1N_2m]^A_n$~\cite{Rost1997}, and the molecular quantum numbers $(n_\lambda n_\mu)^A$~\cite{Watanabe1986,Rost1988}. In the present work we will be mostly concerned with the first few {$^1$S$^e$} and {$^1$P$^o$} states below the $N=2$ threshold, for which the original configuration-based Fano's classification is sufficient. When more precise assignments are required, we will use the Stark classification scheme.
The central rectangle in Fig.~\ref{fig:EnergyScheme}, below the $N=2$ threshold, illustrates the energy and angular distribution of the photoelectrons emitted as a result of the decay of the DESs that form the localised metastable wavepacket obtained by exposing the ground state of the atom to the XUV-pump IR-probe sequence specified above. The spectrum is obtained by letting the fore-front free-electron component of the  wavepacket generated to be absorbed by the box boundaries. The picture shows which DESs are populated most in the process, whose symmetry is visible in the nodal structure of the photoelectron distribution: isotropic for S states, with one nodes at $\theta=90^\circ$ for P states, with two nodes, at the positive and negative magic angle for the D states, and so on. The symmetry, classification, position and width of the first nine resonances is indicated on the right. Three DESs clearly dominate the spectrum in the current conditions: the {$^1$P$^o$} $sp_2^+$ and $sp_3^+$ ($[010]_2^+$ and $[010]_3^+$ in Stark notation) states, and the {$^1$S$^e$} $2p^2$ state ($[100]_2^+$ in Stark notation).
Table~\ref{tab:DESParams} lists the first few DESs converging to the N=2 threshold which are most relevant to the present investigation, and compares the position and width with reference values from the literature.
\begin{table}
\caption{\label{tab:DESParams}Classification and parameters (energy and width) of the first 20 doubly excited states in helium, with natural parity and $L\leq 3$, converging to the $N=2$ threshold. The values in the top row of each group are computed in the present work. The values underneath are taken from the literature. The notation [-n] is an abbreviation for $\times\,10^{-n}$.}
\begin{ruledtabular}
\begin{tabular}{cccccl}
{$\scriptstyle^{2S+1}L^\pi$} & $\scriptstyle_N(K,T)_n^A$& $\scriptstyle[N_1N_2m]_n^A$&conf&$E$~(a.u.)& $\Gamma$~(a.u.)\\
\hline
  {$^1$S$^e$} &  $_2(1,0)^+_2$  &  $[010]^+_2$ & $2s^2$     & -0.777\,863 &  4.54 [-3]  \\
                     &                            &                       &                 & -0.777\,868 & 4.54 [-3]~\cite{Burgers1995,Lindroth1994}\\
  {$^1$D$^e$} &  $_2(1,0)^+_2$ &  $[010]^+_2$ & $2p^2$     &-0.701\,885 &  2.37 [-3]  \\
                       &                           &                       &                 & -0.701\,946 & 2.36 [-3]~\cite{Lindroth1994}\\
  {$^1$P$^o$} &  $_2(1,0)^-_3$   &  $[010]^-_3$  & $sp_2^+$ & -0.693\,061 &  1.37 [-3]  \\
                     &                             &                       &                 & -0.693\,135 & 1.37 [-3]~\cite{Rost1997,Lindroth1994}\\
  {$^1$S$^e$} &  $_2(-1,0)^+_2$ &  $[100]^+_2$ & $2p^2$     & -0.621\,744 &  2.18 [-4]  \\
                     &                             &                       &                 & -0.621\,927 & 2.15 [-4]~\cite{Burgers1995}\\
                     &                             &                       &                 & -0.621\,926 & 2.16 [-4]~\cite{Lindroth1994}\\
  {$^1$P$^o$} &  $_2(0,1)^+_2$  &  $[001]^+_2$ & $sp_3^-$  & -0.597\,074 &  3.84 [-6]  \\
                     &                             &                       &                 & -0.597\,073 & 3.84 [-6]~\cite{Rost1997}\\
  {$^1$S$^e$} &  $_2(1,0)^+_3$  &  $[010]^+_3$ &                  & -0.589\,891 &  1.36 [-3]  \\
                     &                             &                       &                 & -0.589\,895 & 1.36 [-3]~\cite{Burgers1995}\\
                     &                             &                       &                 & -0.589\,89  & 1.36 [-3]~\cite{Lindroth1994}\\
  {$^1$D$^e$} &  $_2(1,0)^+_3$  &  $[010]^+_3$ &                  & -0.569\,205 &  5.58 [-4]  \\
                     &                             &                       &                 & -0.569\,22 & 5.56 [-4]~\cite{Lindroth1994}\\
  {$^1$P$^o$} &  $_2(1,0)^-_4$   &  $[010]^-_4$  & $sp_3^+$ & -0.564\,072 &  3.01 [-4]  \\
                     &                             &                       &                 & -0.564\,085 & 3.01 [-4]~\cite{Rost1997}\\
  {$^1$F$^o$} &  $_2(1,0)^-_3$   &  $[010]^-_3$  & & -0.558\,265 &  1.30 [-5]  \\
                     &                             &                       & & -0.558\,28 & 1.28 [-5]~\cite{Lindroth1994}\\
  {$^1$D$^e$} &  $_2(0,1)^-_3$   &  $[001]^-_3$  & & -0.556\,420 &  2.00 [-5]  \\
                     &                             &                       & & -0.556\,43   & 2.00 [-5]~\cite{Lindroth1994}\\
  {$^1$S$^e$} &  $_2(-1,0)^+_3$ &  $[100]^+_3$ & & -0.548\,065 &  7.58 [-5]  \\
                     &                             &                       & & -0.548\,086 & 7.48 [-5]~\cite{Burgers1995}\\
  {$^1$P$^o$} &  $_2(0,1)^+_3$  &  $[001]^+_3$ & & -0.546\,489 &  2.01 [-6]  \\
                     &                             &                       & & -0.546\,493 & 2.02 [-6]~\cite{Rost1997}\\
  {$^1$S$^e$} &  $_2(1,0)^+_4$  &  $[010]^+_4$ & & -0.544\,879 &  4.92 [-4]  \\
                     &                             &                       & & -0.544\,882 & 4.92 [-4]~\cite{Burgers1995}\\
  {$^1$D$^e$} &  $_2(1,0)^+_4$  &  $[010]^+_4$ & & -0.536\,719 &  2.33 [-4]  \\
  {$^1$P$^o$} &  $_2(1,0)^-_5$   &  $[010]^-_5$  & & -0.534\,356 &  1.28 [-4]  \\
                     &                             &                       & & -0.534\,363 & 1.28 [-4]~\cite{Rost1997}\\
  {$^1$F$^o$} &  $_2(1,0)^-_4$   &  $[010]^-_4$  & & -0.532\,246 &  7.24 [-6]  \\
  {$^1$D$^e$} &  $_2(0,1)^-_4$   &  $[001]^-_3$  & & -0.531\,506 &  1.11 [-5]  \\
  {$^1$S$^e$} &  $_2(-1,0)^+_4$ &  $[100]^+_4$ & & -0.527\,704 &  4.72 [-5]  \\
                     &                             &                       & & -0.527\,717 & 4.62 [-5]~\cite{Burgers1995}\\
  {$^1$P$^o$} &  $_2(0,1)^+_4$  &  $[001]^+_4$ & & -0.527\,291 &  9.79 [-7]  \\
                     &                             &                       & & -0.527\,298 & 9.82 [-7]~\cite{Rost1997}\\
  {$^1$S$^e$} &  $_2(1,0)^+_5$  &  $[010]^+_5$ & & -0.526\,682 &  2.18 [-4]  \\
                     &                             &                       & & -0.526\,687 & 2.18 [-4]~\cite{Burgers1995}\\
\end{tabular}
\end{ruledtabular}
\end{table}
\begin{figure*}[hbtp!]
\includegraphics[width=\linewidth]{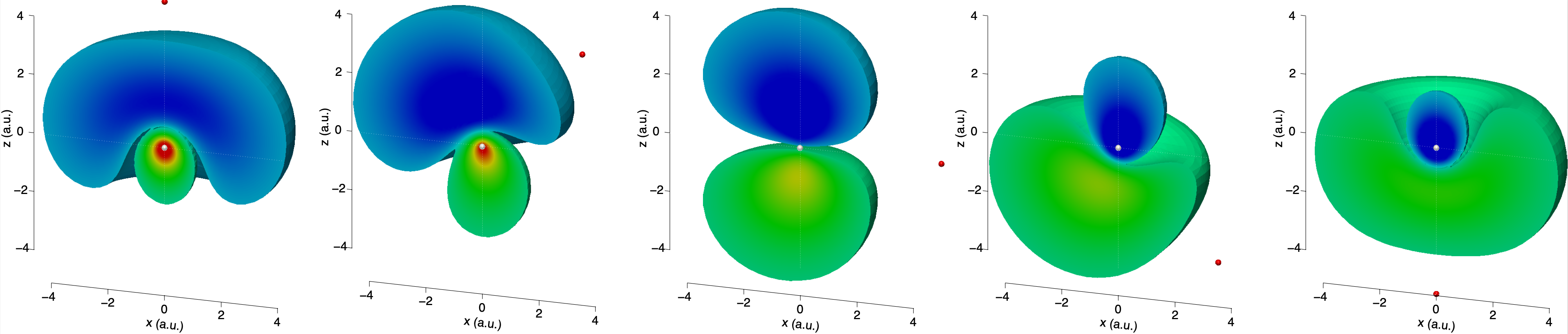}
\caption{\label{fig:StaticCorrelation} Normalised {$^1$P$^o$} component of the two-electron wave packet, generated by the interaction of the ground state of the helium atom with the XUV pulse, for selected fixed positions in space of one of the two electrons (indicated by a red solid sphere) as a function of the position of the other electron. The white sphere at the origin indicates the position of the nucleus. The electron wavepacket is colored coded in proportion to the magnitude of its real component, to highlight the change in sign across the node, and the its modulation through space.}
\end{figure*}

If electronic correlation affects the properties of many bound states of atoms and molecules~\cite{Hattig2011}, it completely dominates those of doubly excited states~\cite{Tanner2000}. Figure~\ref{fig:StaticCorrelation} shows the conditional electron probability density $P(\vec{r}_2|\vec{r}_1)$ of the {$^{1}$P$^o$} component of the wave function excited by the attosecond XUV pulse, for several positions of one of the two electrons at a fixed distance of slightly more than $4$~au from the nucleus.  The distribution has been cut at a threshold value of the conditional density. The wave packet has a distinct $sp^+$ character: the parent ion is in the Stark state polarised towards the outer electron. The parent-ion wave function entangled to the outer electron at fixed positions in space reproduces a polarized hydrogenic Stark state~\cite{Stodolna2013}. On top of such an exquisitely static-polarisation effect, inter-electronic repulsion is evident in the way the parent ion is distorted away from the outer electron. Such long range correlation has an adiabatic character that explains the energetics of the metastable states, rather than its Auger decay.

\begin{figure}[hbtp!]
\includegraphics[width=\linewidth]{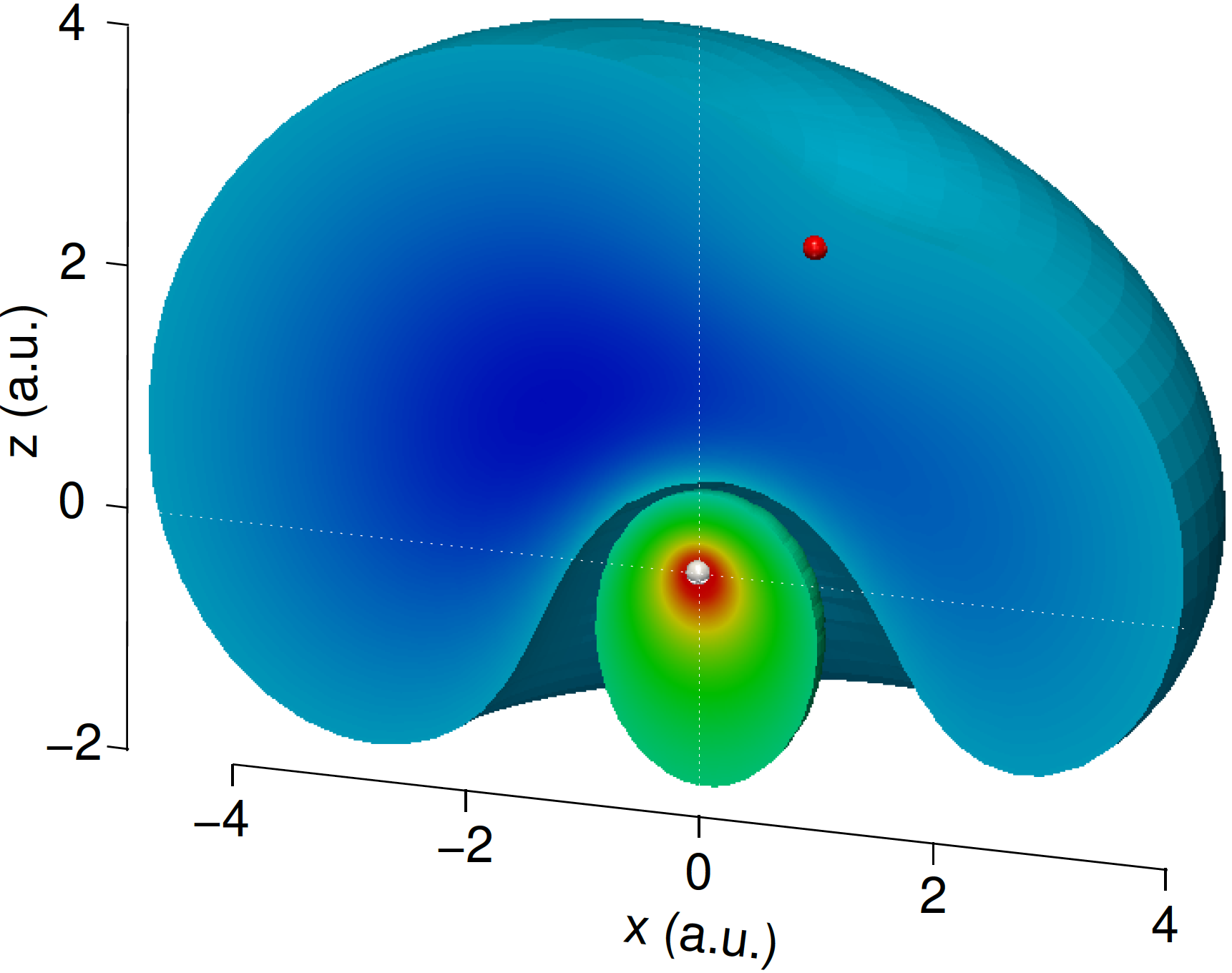}
\caption{\label{fig:CoulombHole} Metastable wave packet (cmp with fig.~\ref{fig:StaticCorrelation}) with the position of one of the electrons (red sphere)  penetrating the region of the Stark parent-ion state. The depletion of electron density (Coulomb hole) around the fixed electron is at the origin of the Auger decay of the doubly excited states, which acquire a finite lifetime as a result.}
\end{figure}

Figure~\ref{fig:CoulombHole} shows the conditional density when one of the two electrons is at a shorter distance, within the threshold set for the density. In this case, the electron density is depleted in proximity of the reference electron (Coulomb hole)~\cite{Hattig2011}. This dynamical correlation allows the two electrons to exchange a large amount of energy, leading to the eventual collapse of one of them to the $1s$ ground state of He$^+$ and to the ejection of the other with an asymptotic energy as large as $40$~eV.

\begin{figure}[hbtp!]
\includegraphics[width=\linewidth]{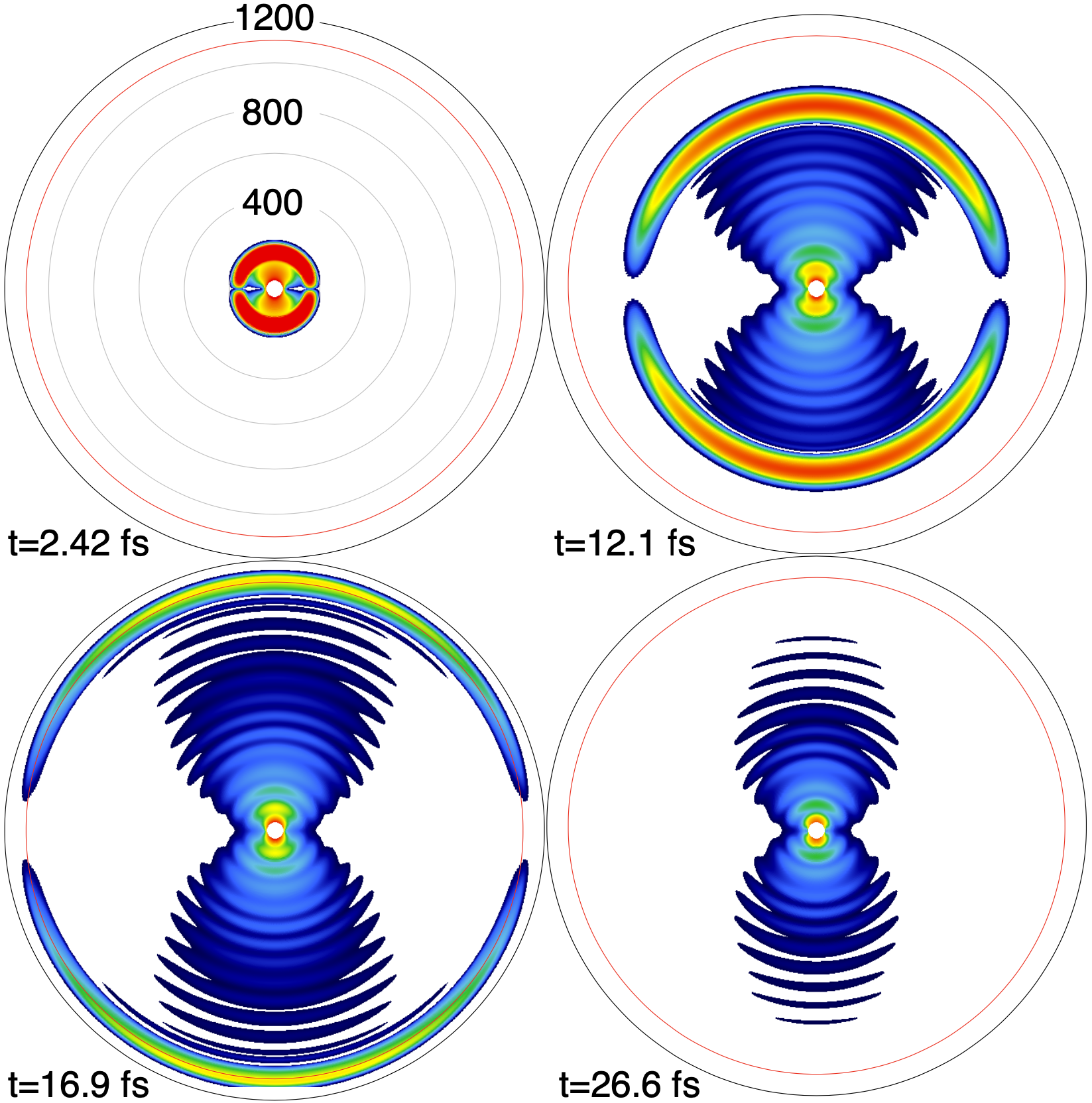}
\caption{\label{fig:RhovsT} Propagation in the box of the electron density generated by the XUV pulse (color is on a log scale). The first wavefront corresponds to the direct-ionization component of the wave packet. The trailing maxima are predominantly due to the beating between the photoelectron amplitude from the decay of the $sp_2^+$ and $sp_3^+$ DESs. The wave packet disappears once it enters the CAP region (red line). }
\end{figure}

\begin{figure}[hbtp!]
\includegraphics[width=\linewidth]{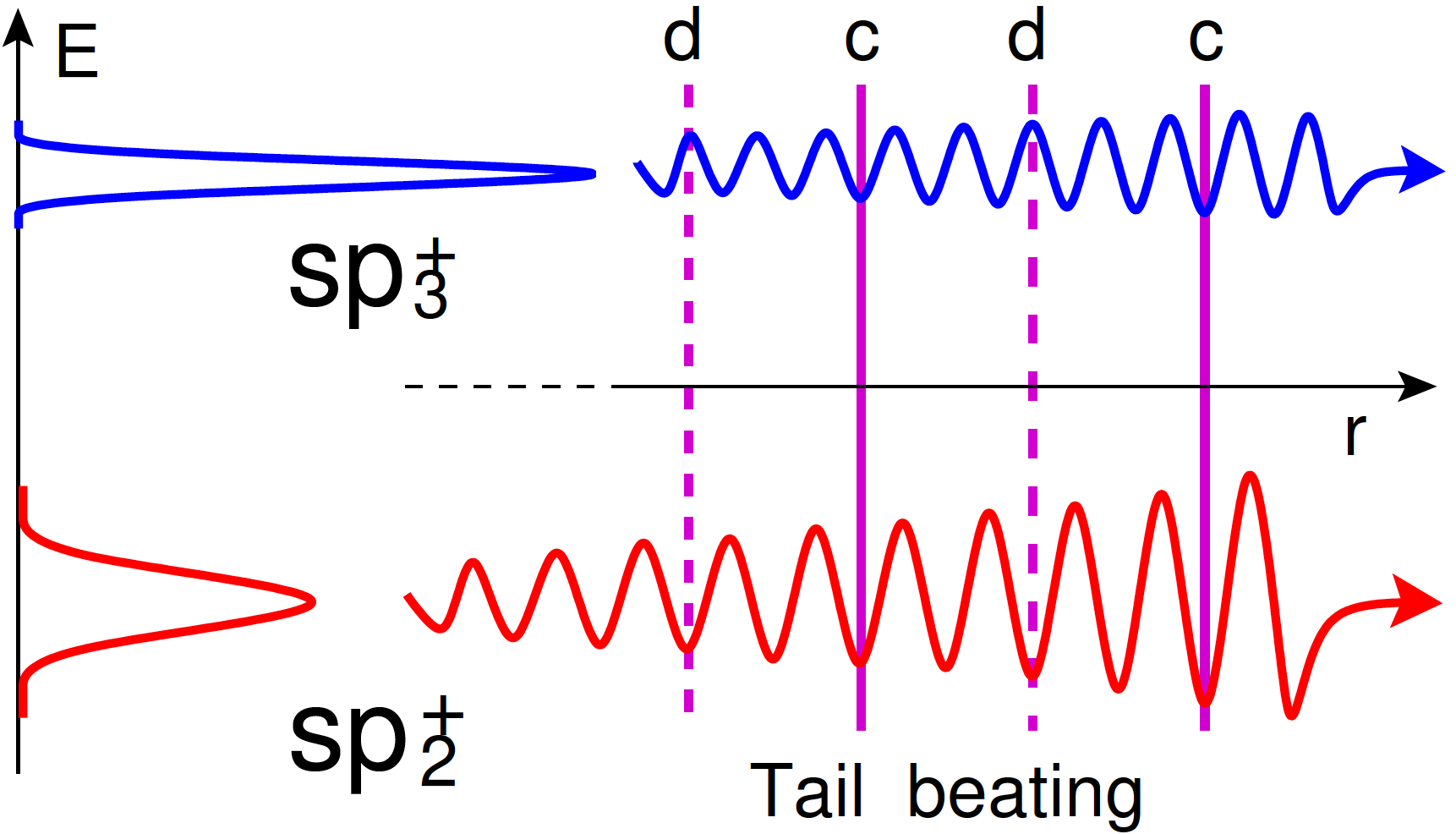}
\caption{\label{fig:WaveInterference} The interference between the Auger decay amplitudes from the $sp_2^+$ and $sp_3^+$ autoionizing states gives rise to a beating in the photoelectron density as a function of the distance from the parent ion. The picture shows schematically the real part of the resonances' tails, and how they alternatively interfere constructively (c) and destructively (d) with a radial periodicity of $2\pi/(k_{sp_3^+}-k_{sp_2^+})$, where $k=\sqrt{2\epsilon}$ is the peak value of the Auger electron's momentum.}
\end{figure}

Figure~\ref{fig:RhovsT} shows the aspect of the electron density in the quantization box, up to a radial distance of 1200~a.u., at four different times after the excitation event. At $t=12.1$~fs, the wavepacket has reached a radius of about $800$~a.u., and it is possible to recognize some characteristic features. At the largest distance, photoelectrons from the one-photon direct-ionization process form a first prominent wavefront, moving outward with group velocity $\sqrt{2(\omega_\XUV-I.P.)}$. At short radii, there is a non-negligible probability for both electrons to reside (R $\leq$ 50~a.u., not represented in the picture, where it is recognizable as a white circular spot). 
This is where electronic correlation exerts most of its influence. In particular, it is in this central region that metastable states are located, and decay. The region between the short-range region and the outermost direct-ionization wavefront is dominated by a series of outgoing wavefronts. These wavefronts are due to the interference between the autoionization amplitude originating from the concurrent Auger decay of multiple {$^1$P$^o$} transiently bound states. As the intensity of the spectral lines in Fig.~\ref{fig:EnergyScheme} indicate, the two dominant components, at least at short time, come from the $sp_n^+$ states ($n=2$, $3$). Each of them gives rise to a {$^1$P$^o$} Siegert state whose asymptotic spatial part ($r_2\gg 1$) has the approximate form
\begin{equation}
\varphi_{n}^{\mathrm{Si}}(\vec{r}_1,\vec{r}_2)\propto \mathcal{S}\,\phi_{1s}(\vec{r}_1)\,\frac{Y_{L0}(\hat{r}_2)}{r_2}\,\exp\left[\frac{\Gamma_n r_2}{2k_n}+ik_n r_2\right],
\end{equation}
where $\mathcal{S}=2^{-1/2}(1+\mathcal{P}_{12})$ is the symmetrizer and $\Gamma_n$ is the resonance width. Indeed, given $E_n=\bar{E}_n-i\Gamma_n/2 = (k_n^2-ik^{Im}_{n})/2$ is the complex energy of the resonance with respect to the ionization threshold, and the assumption that the resonance is narrow compared to the energy of the photoelectron it releases, $\Gamma_n\ll \bar{E}_n$, it follows that $k_n\simeq \sqrt{2\bar{E}_n}$ and $k_n^{Im}\simeq \Gamma_n/(2k_n)$. The principle behind the spatial interference between the Auger amplitude from the two concurrently decayed resonances is illustrated in Fig.~\ref{fig:WaveInterference}. 
Since the two DES emit the photoelectron with a different radial momentum, the reduced radial part of the Auger electron wavefunction $u(r,t)$, in the intermediate region, has the approximate form
\begin{equation}
u(r,t)\propto \sum_{n=2,3}c_n\, \exp\left[\frac{\Gamma_n}{2k_n} (r-k_nt)+i(k_nr-\bar{E}_n t)\right],
\end{equation}
where $c_2$ and $c_3$ are fixed coefficients. The oscillating part of the radial Auger density, therefore, is proportional to
\begin{equation}
|u(r,t)|^2\propto |c_2c_3| e^{\bar{K} r-\bar{\Gamma}t}\cos\left\{\Delta k [r-v_g (t-\tau_{\textsc{AI}})]\right\},
\end{equation}
where $\bar{K}=(\Gamma_2/k_2+\Gamma_3/k_3)/2$, $\bar{\Gamma}=(\Gamma_2+\Gamma_3)/2$, $\Delta k = k_3 - k_2$, and $v_g=(\bar{E}_3-\bar{E}_2)/\Delta k$ is the radial group velocity. The quantity $\tau_{\mathrm{Coll}}=\arg(c_3/c_2)/(\bar{E}_3-\bar{E}_2)$, which we may call collisional delay (see below for more details), is the time it takes, from the initial excitation of the two resonances at $t=0$, to emit the photoelectron in phase, thus leading to the formation of a photoemission wavefront. The four panels in figure~\ref{fig:RhovsT} illustrate this dynamics, as well as the dissipative effect of the complex absorption potential in the outer 100~a.u.-thick layer of the quantization box.

Due to the dispersive character of electron propagation, the profile of the radial photoelectron distribution eventually converges to the asymptotic energy distribution,
\begin{equation}
\frac{dP}{dE}=\frac{1}{2E}\lim_{t\to\infty} \frac{dP(r;t)}{d\ln(r)}.
\end{equation} 
At a microscopic distance from the atom, it is nevertheless possible, at least in principle, to detect the peaked electron emission, as a function of time. A similar experiment has in fact been realized with rubidium Rydberg wavepackets in an external dc field~\cite{Lankhuijzen1996}. In this context, it is interesting to point out how the direct-photoemission wavefront is separated from the trailing Auger amplitude by a deep minimum, which is visible in Fig.~\ref{fig:RhovsT} as a narrow white crevice for $t=12.1$~fs and $t=16.9$~fs. This deep minimum is due to the destructive interference between the direct photoemission and the resonant decay. At long times, this minimum is preserved and it maps to the zero of the Fano profile of the $sp_2^+$ doubly excited state.

\begin{figure*}[hbtp!]
\includegraphics[width=\linewidth]{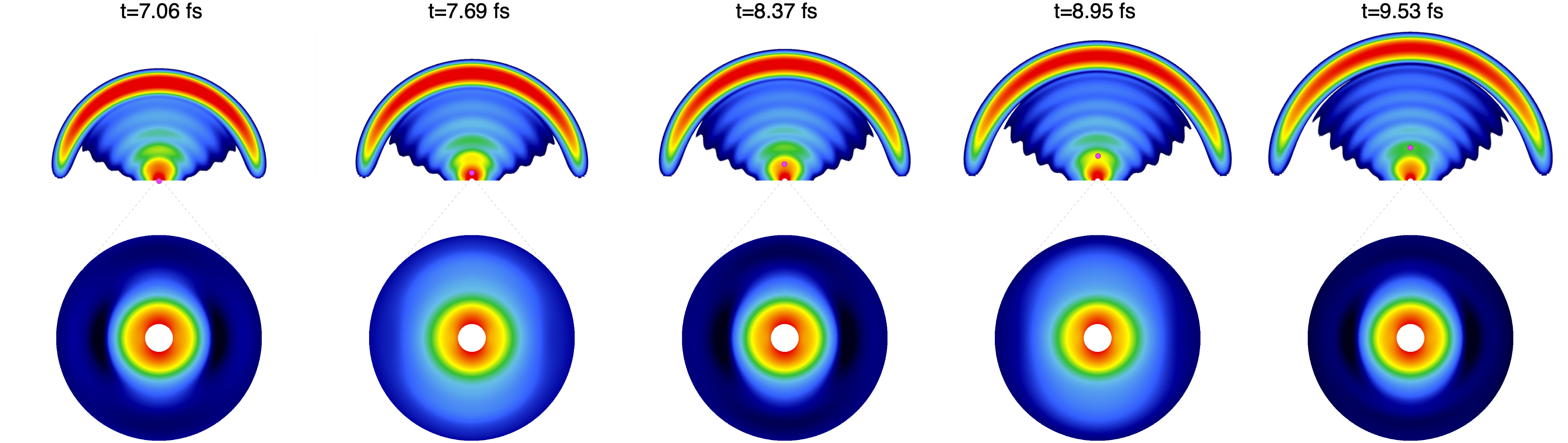}
\caption{\label{fig:CollisionalInterpretation} The periodicity of the photoelectron wavefront reproduce the periodicity of the concerted breathing motion of the two electrons in the metastable wavepacket, which has a predominant $sp_n^+$ character~\cite{Argenti2010}. A new wavefront is emitted whenever the metastable wavepacket reaches maximum contraction, as in these conditions the two electrons are in closest proximity to each other and hence the rate of energy exchange between them is highest.}
\end{figure*}

Figure~\ref{fig:CollisionalInterpretation} correlates the density of the Auger electron at large distance (top row), up to 800 Bohr radii, with the density at short range (bottom row), within just 15 Bohr radii from the nucleus, computed considering only the $^1$P$^o$ component of the wave packet, at five consecutive times after the excitation XUV pulse. Setting to zero the population of the ground state, which does not contribute to the ionization dynamics once the external field is over, greatly clarifies the short-range electron dynamics, since the ground state would completely obscure the excited wave packet, due to its disproportionately larger population. 
Close to the parent ion, apart for the ground state, the density is dominated: i) by the population of the $1s$ He$^+$ parent ion entangled with the fast outgoing photoelectrons; ii) to a smaller extent, by the population of the $2s$ and $2p$ parent ions entangled with the slow outgoing photoelectrons in the $N=2$ channels; iii) by the density of the two-electron metastable wave packet formed by the doubly excited states. The coherent population of the $sp_2^+$ and $sp_3^+$ states, in particular, gives rise at short range to a quasi periodic concerted motion of the two electrons. Indeed, this motion was accurately reconstructed from attosecond transient absorption measurements in~\cite{OttNature2014}, where it was found to be in excellent agreement with the~\emph{ab initio} optical observables obtained from simulations used the very same suite of programs described in this work. As explained in~\cite{Argenti2010}, furthermore, the contraction of the metastable wavepacket coincides with the emission of a photoelectron wavefront. This correspondence is schematically illustrated in Fig.~\ref{fig:CollisionalInterpretation}. The Auger electron density periodically exhibits a broad wavefront that peaks near the origin, as seen in the long-range plot (see top row for $t=7.06$~fs, $t=8.37$~fs, and $t=9.53$~fs). At the same time, the metastable wavepacket is the most contracted, at short range (bottom row). This correspondence in the ejection of the photoelectrons associated to the Auger decay of the metastable wave packet points to the classical character of the electron-electron collision process that underpins the large exchange of energy required for one electron to collapse to the $1s$ He$^+$ state while the other is ejected to the continuum. A similar phenomenon was observed with pairs of colliding Rydberg wavepackets~\cite{Pisharody2004} in doubly excited barium atom and in doubly excited states in Magnesium~\cite{Warntjes1999}. In the latter case, one of the two electrons has $n=3$, while the other is excited to a superposition of Rydberg states with very high principal quantum numbers. It was found that the atom decays only when the Rydberg wave packet returns in proximity of the nucleus, where it can exchange energy with the $n=3$ parent ion. 

The trajectory of the first peak at $t=7.06$~fs is highlighted in the frames by a purple circle, which moves out with constant speed $v_g$. Based on the energy difference between the $sp_2^+$ and the $sp_3^+$ states, $\Delta E \simeq 3.51$~eV, we expect a beating with a period $T\simeq$1.18~fs, which is in line with the average between the first and last peak, of $1.23$~fs (the periodicity is not exact, due to the small interference from higher terms in the resonance series). What we call here collisional delay $\tau_{\mathrm{Coll}}$ is also the time at which the two electrons in the metastable wavepacket get to first be at their closest, after the initial excitation, thus maximizing the chance of a collision that results in the Auger emission (hence the name). By attempting a minimum-square fit of the three maximum contraction times $t=7.06$~fs, $t=8.37$~fs, and $t=9.53$~fs with the formula $t_n=\tau_{\mathrm{Col}}+nT$, the best estimate for the collision time is $\tau_{\mathrm{Col}}=0.06~$fs$~\pm0.49$~fs. This result is obviously compatible with $\tau_{\mathrm{Col}}=0$~fs, i.e., with the Auger decay rate peaking at $t=0$ and at all subsequent multiples of $T$. Indeed, from the initial $1s^2$ state, which is the most compact bound state of helium, the XUV pulse gives rise to a scattering wave packet with spatial component approximately proportional to $(z_1+z_2)1s(r_1)1s(r_2)$, itself still more contracted than either the $sp_2^+$ or the $sp_3^+$ states. It is perfectly reasonable, therefore, that for this system the first time of closest encounter is also the time at which the wave packet is generated. 

In general, the XUV pump pulse can excite more than two resonances. In fact, it can excite at once the whole autoionizing $N=2$ Rydberg series. Since the lifetime increases along the series ($\Gamma_n\sim 1/n^3$), and the probability with which the resonances are excited is itself proportional to $\Gamma$, the shortest-lived resonances dominate the decay dynamics at first, to be subsequently supplanted by the other resonances. The emission of Auger wavefronts, therefore, becomes irregular, negatively chirped, and fainter, as the metastable wavepacket decays.

\subsection{Time Evolution in the Small Basis\label{sec:result_smallBasis}}

The ionization yields published in~\cite{Argenti2010} were obtained with simulations in a quantization box with a radius of 400~a.u. In such a small box, already for time delays of few femtoseconds (XUV first), the photoelectron starts to be absorbed at the box boundary well before the end of the IR probe pulse. 
\begin{figure}[hbtp!]
\includegraphics[width=\linewidth]{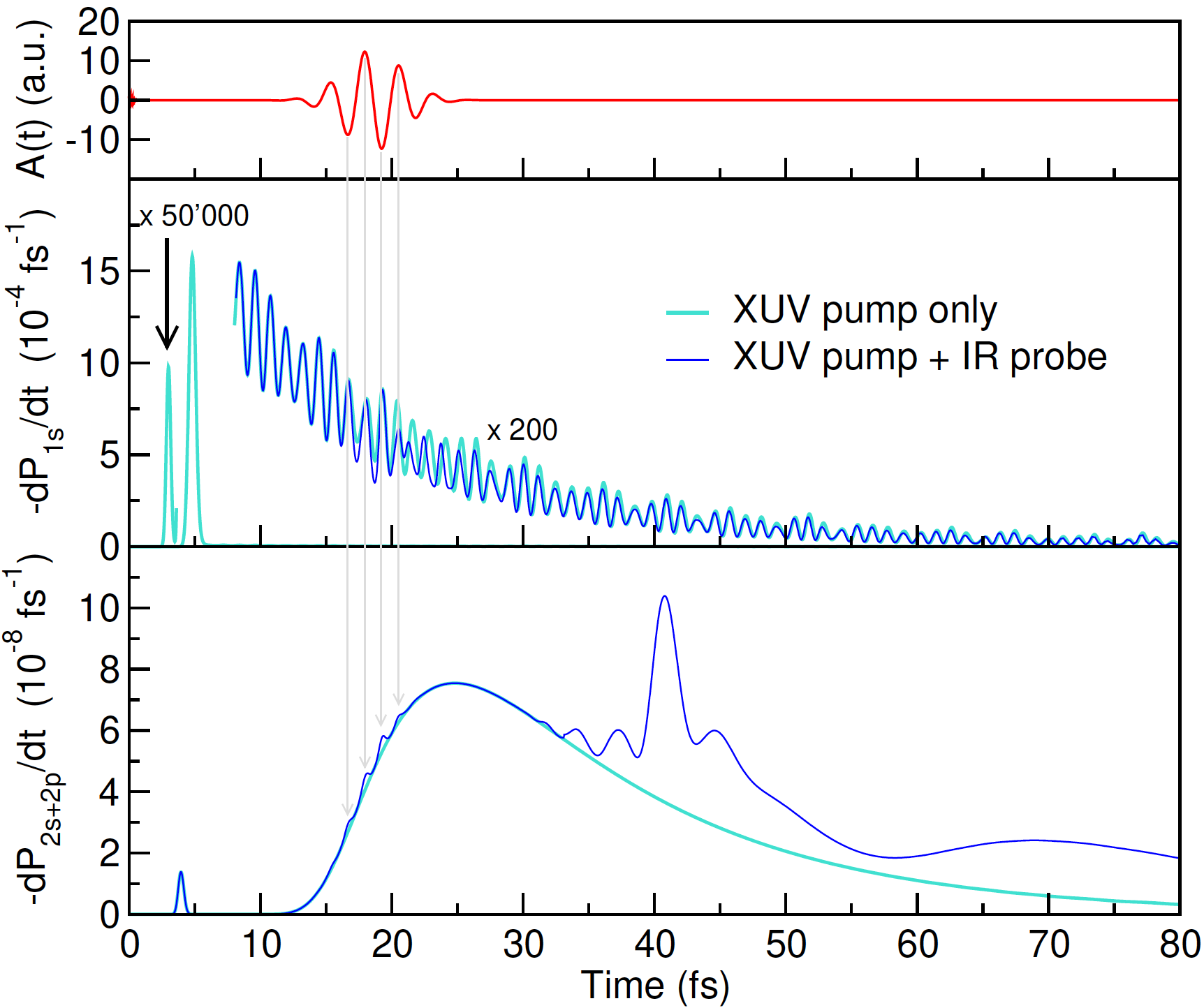}
\caption{\label{fig:Qrate} Absorption rate of the photoelectrons in the $1s$ channel (mid panel) and $N=2$ ($2s+2p$) channels (bottom panel), as a function of time, for the simulation with a small basis in a box with 400~au radius. Top panel: amplitude of the total vector potential. Mid panel: most of the signal comes from a peak that reaches the absorber 5~fs after the XUV pulse. This peak corresponds to the direct-ionization signal. A signal five orders of magnitude smaller reaches the detector after barely 3~fs. This is the signal due to the absorption of two photons. Finally, at large time, a signal two orders of magnitude smaller accounts for the slow trailing decay of the doubly excited states, which in the present conditions account for about 10\% of the excitation probability.}
\end{figure}
Figure~\ref{fig:Qrate} shows the absorption rate at the boundary in the N=1 ($1s\varepsilon_\ell$) and N=2 ($2\ell\varepsilon_{\ell'}$) channels, as a function of time, in the case of an XUV-pump IR-probe delay of approximately 18.6~fs. It is clear that, in these conditions, a significant fraction of the electrons in the $N=2$ channel have already been absorbed by the CAPs at the end of the pulse. This means that the partial ionization yield cannot be computed by integrating the photoelectron spectrum at the end of the simulation, as the overall wavefunction is already compromised at that time. In the $1s$ channel, we clearly see the signal of the direct ionisation, followed by the beating between the several resonances that are populated by the pulse. If the resonances are not populated, then the tails disappear entirely. In the N=2 channels, which are populated at arbitrarily low photoelectron energies, we notice a very small signal at short times. This is the signal corresponding to the absorption of two photons.

One possible way to compute the yield in the $2s$ and $2p$ channel from these data is to integrate the absorption rate up to very large time, until convergence is reached. Unfortunately, this approach is largely compromised by the interference of the doubly-excited states with principal quantum number $n\geq12$ that survive the IR pulse. These states form a coherent wavepacket that reaches the CAPs region, where it is absorbed, thus causing an overestimate of the shake-up yield. This interference cannot be disentangled from the signal from the electrons above the threshold, as high-lying Rydberg satellites and low-energy shake-up electrons reach the boundary in the same time range. Furthermore, notice the several small peaks featured in the $N=2$ absorption rate, in correspondence of the maximum of the vector potential. These peaks, which are not physical, are due to the instantaneous polarization, in velocity gauge, of the $1s$ orbital, which acquires a small $2p$ component. As a consequence, the outgoing electrons in the $1s$ channel (which, at the time of the probe pulse in the figure, originates from the decay of the DESs) temporarily acquires a small $2p$-channel character at the peak of the IR field, thus resulting in a fictitious contribution to the $N=2$ channel. 

In~\cite{Argenti2010} these difficulties were circumvented as follows. A time close to the end of the IR pulse, $t=1200$~a.u., before the photoelectron generated in the $N=2$ channel by the IR could reach the absorption boundaries, was selected. The population of the $2s$ and $2p$ channels was evaluated by adding to the population that had already been absorbed the residual population computed by projecting the wavepacket still in the box on a complete set of $|\psi_{2\ell\varepsilon_{\ell'}}^{\Gamma(-)}\rangle$ scattering states,
\begin{equation}
P_{2\ell} = \int_0^t dt' \frac{dP_{2\ell}(t')}{dt'} + \sum_{\Gamma\ell'}\int_0^\infty d\varepsilon \left|\langle \psi_{2\ell\varepsilon_{\ell'}}^{\Gamma(-)}|\Psi(t)\right|^2.
\end{equation}

In this section we compare the results obtained with this original method with those obtained in an equivalent basis but in a quantization box three times as large, $R_{\mathrm{max}}=1200$~a.u., where, by the end of the external pulses, none of the photoelectrons in the $N=2$ channels has yet reached the boundary. In the larger box, therefore, the population can be consistently computed by integrating the photoelectron spectrum,
\begin{equation}
P_{2\ell} = \sum_{\Gamma\ell'}\int_0^{\infty} d\varepsilon \left|\langle \psi_{2\ell\varepsilon_{\ell'}}^{\Gamma(-)} | \Psi(t_{\mathrm{f}})\rangle \right|^2
\end{equation}
The old yields in absence of IR pulse were $P_{2s} = 5.400[-7]$ and  $P_{2p} = 1.457[-6]$, to be compared with the newly computed yields, in absence of the IR pulse, of $P_{2s} = 4.788[-7]$ and $P_{2p} = 1.457[-6]$, which demonstrate that the original procedure was already quite accurate.

\begin{figure}[hbtp!]
\includegraphics[width=\columnwidth]{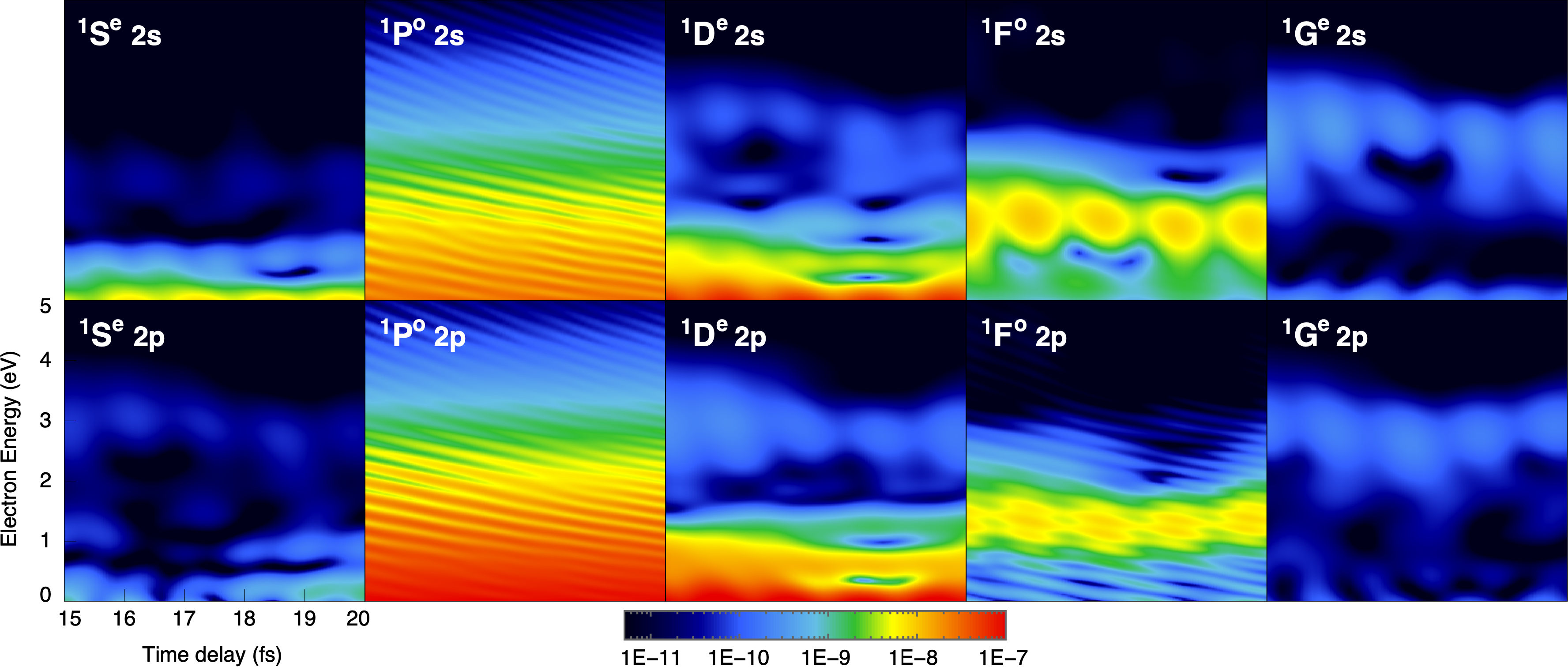}
\caption{\label{fig:dPaGdEvsT_unc} Symmetry-resolved partial differential photoelectron distributions in the $2s$ and $2p$ channels as a function of the time delay and of the photoelectron energy. The population for $L>4$ is negligible.}
\end{figure}

Figure~\ref{fig:dPaGdEvsT_unc} shows the energy- and symmetry-resolved photoelectron distribution for the $2s$ and $2p$ channels, as a function of the time delay, in the delay interval between 15 and 20~fs, from which the phase of the oscillation of the partial yield was originally reconstructed. 
In absence of the IR probe, of course, only the {$^1$P$^o$} symmetry is populated. In the presence of the IR, additional transitions become possible. In particular, the IR promotes the ATI ionization of the {$^1$P$^o$} DESs above the N=2 threshold. The interference between the featureless direct-ionization amplitude in the $2\ell$ {$^1$P$^o$} channels and the indirect ATI amplitude to the same symmetry gives rise to the well known hyperbolic holographic fringes that are well visible in the {$^1$P$^o$} symmetry. Most of the other symmetries receive contributions exclusively from the ATI DESs ionization amplitudes. Therefore, they show the characteristic beatings due to the dynamics of the DES metastable wavepacket, but they do not exhibit the interference fringes. The {$^1$F$^o$} symmetry in the $2p$ channel is a notable exception. In this case we do see clear hyperbolic fringes. The reason is that a secondary effect of the IR is to dress the 2p parent ion. The effect of such dressing is to bring slightly out of phase the $p_z$ component, compared to the $p_x$ and $p_y$ components. As a result, part of the original $[2p\varepsilon_{d}]_{^1\mathrm{P}^o}$ channel, which does exhibit interference fringes, is converted to the $[2p\varepsilon_{d}]_{^1\mathrm{F}^o}$ channel. The peculiar interference fringes in the latter, therefore, are not a numerical artifact.

\begin{figure}[hbtp!]
\includegraphics[width=\linewidth]{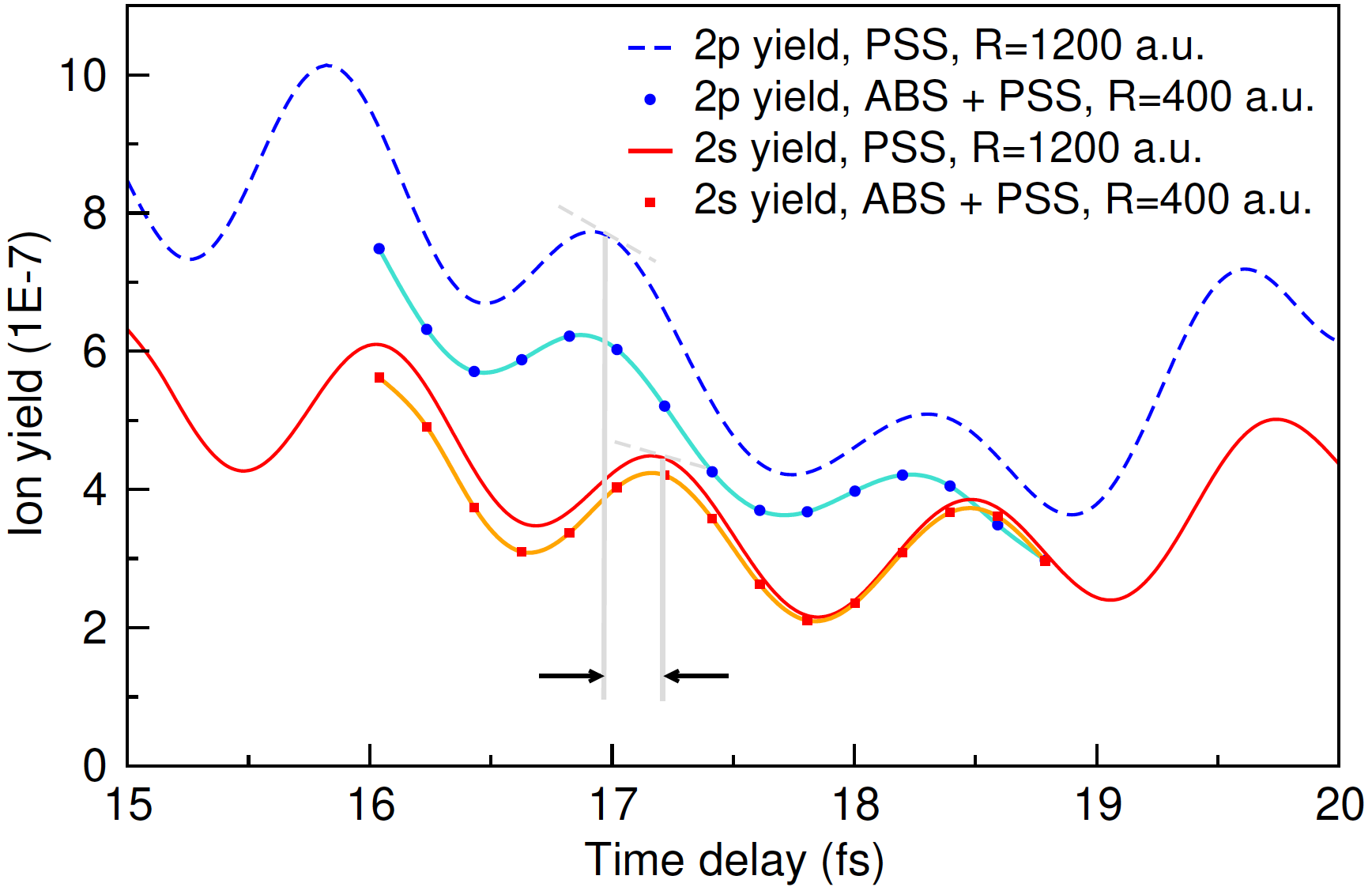}
\caption{\label{fig:PaGvsT_unc} Partial yield of the $2s$ and $2p$ He$^+$ parent ions as a function of the pump-probe time delay.  The  bullets, with interpolating curves to guide the eye, are the data computed in~\cite{Argenti2010}, in a box with a radius of only 400~a.u., by mixing different methods to count the partial ionization probability. The extended curves, on the other hand, are the newly calculated data, in a box with radius of 1200~a.u., and with a more robust numerical method to disentangle the asymptotic partial ionization probability. Apart a background shift, the two data are in essential agreement, and in particular so is the phase shift between the beating in the $2s$ and the $2p$ channels.}
\end{figure}

Figure~\ref{fig:PaGvsT_unc} compares the integral $2s$ and $2p$ yields computed in the 400~a.u. box for~\cite{Argenti2010} with yields computed in the larger 1200~a.u. box, still in the minimal CC basis. Overall, the old $2p$ yield was slightly underestimated, but one of the main conclusion of the original paper, namely, that the two yields oscillate with a distinct phase offset still stand. It is already visible to the naked eye that the two phase offsets are in good agreement with each other. 
In the following, we will examine a new set of results conducted with a much larger short-range configuration-interaction basis. When electron-electron correlation is fully taken into account, the phase offset between the $2s$ and $2p$ yields is in fact much larger than these original calculations suggested, thus making the contrast highlighted in~\cite{Argenti2010} even more striking.

\subsection{Time Evolution in the Large basis}
In this section, we analyze the prediction of the present method for the same pump-probe process described in the previous section, for a large box, $R_{\mathrm{max}}=1200$~a.u., but where the pseudo-channel is the full-CI space generated by all the short-range B-splines orbitals with angular momentum up to $\ell_{\mathrm{max}}=5$. In contrast to the calculation with a minimal close-coupling basis, this larger basis is expected to yield quantitative gauge agreement. 
\begin{figure}[hbtp!]
\includegraphics[width=\linewidth]{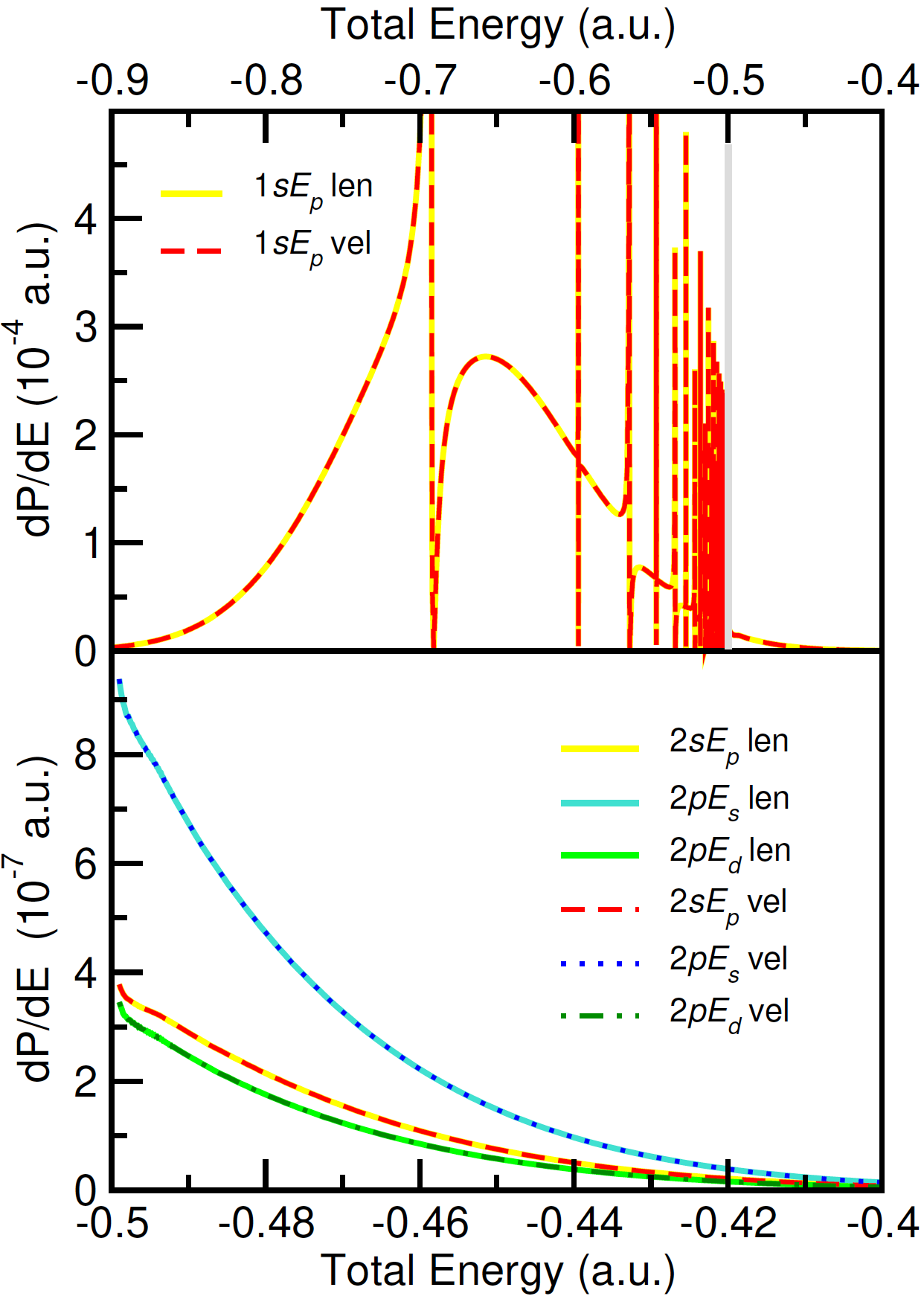}
\caption{\label{fig:dPdE1PoVvsL} Partial photoelectron distribution in {$^1$P$^o$} symmetry after the absorption of one XUV photon computed in both length (continuous lines) and velocity gauge (segmented lines). The vertical gray line indicate the N=2 threshold.}
\end{figure}
Figure~\ref{fig:dPdE1PoVvsL} shows the differential photoelectron spectrum in the $1s\varepsilon_p$, $2s\varepsilon_p$, and $2p\varepsilon_{s/d}$ {$^1$P$^o$} channels, due to the XUV pump pulse alone, computed in both length and velocity gauge. This figure illustrates several aspects of the ionization event with the present pump spectrum in the larger basis. First, the total ionization yield is dominated by the contribution from the $1s$ channel below the $N=2$ threshold, which surpasses the shake-up contribution by three orders of magnitude. Even at total energies above the $N=2$ threshold, the direct-ionization yield is ten times larger than the overall shake-up yield. Conversely, below the $N=2$ channel, the population of electrons transiently bound in DESs immediately after the pulse, while a fraction of the direct-ionization yield, is still substantially larger than the shake-up yield. Second, the large box allows us to accurately describe a large number (more than $20$) resonances for each autoionization series. Third, as anticipated, the two gauges are in excellent agreement in all the channels and for all the energies, which suggests that the description of the {$^1$P$^o$} sector of the configuration space in the energy interval of interest is converged. 

\begin{figure}[hbtp!]
\includegraphics[width=\linewidth]{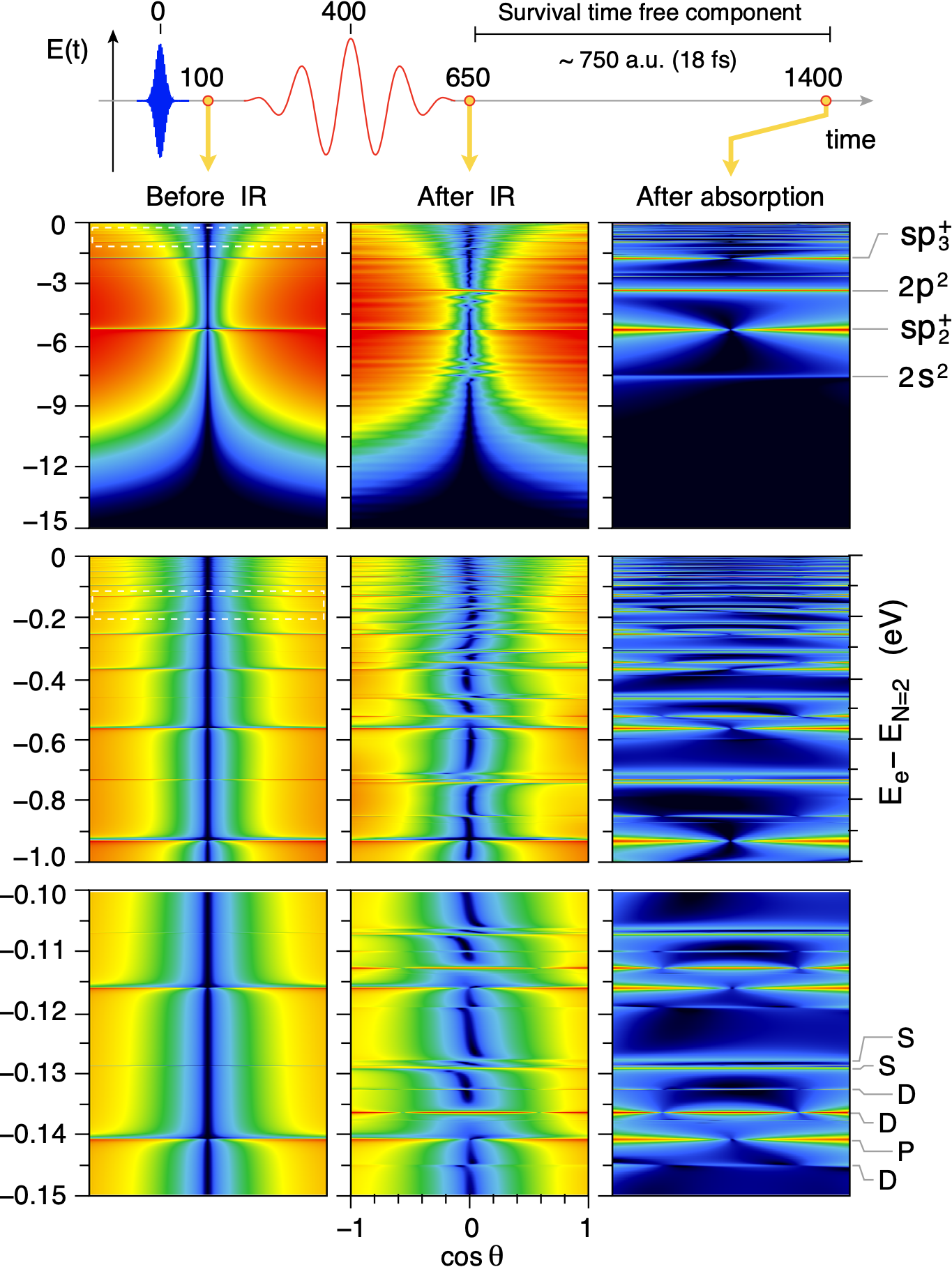}
\caption{\label{fig:TelescopicSeries} 
Resonant photoelectron spectrum in the proximity of the DES below the N=2 threshold, at various times during the propagation (earlier to the left, later to the right), and at various  levels of detail (lower on top, higher on bottom). Left column: before the IR, central column, immediately after the IR, $\tau=$400~a.u., when only the {$^1$P$^o$} states are visible. Central column: immediately after the IR pulse, when the spectrum already coincides with the asymptotic observable. Right column: at $t=1400$~a.u., a time sufficiently large for both the direct-ionization electrons created by the initial XUV pulse and the sideband free electrons created by the action of the VIS probe pulse on the metastable part of the wave packet to reach the box boundary and be absorbed. As a consequence, the plot mostly represent the energy spectrum of the non-decayed localised part of the doubly excited wave packet. }
\end{figure}

Figure~\ref{fig:TelescopicSeries} shows the asymptotic spectrum of the time-dependent wave packet, in $1s$-channel, resolved in energy and photoemission angle, at three characteristic stages of the XUV-pump IR-probe numerical simulation: i) in the field-free interval between pump and probe, ii) immediately after the IR-probe, and iii) shortly after the $1s$ direct-ionization component of the wave function has been absorbed by the box boundaries. The figure also illustrates the remarkable energy resolution that can be achieved by computing the photoelectron spectrum from the projection of the wave packet on the scattering states of the many-body field-free hamiltonian. 
The spectrum before the probe pulse is just a replica of the plot in Fig.~\ref{fig:dPdE1PoVvsL}, since the $1s\varepsilon_{p}$ channel has a trivial $\cos^2\theta$ angular distribution, with a single node at $\theta=90^\circ$.
The interaction with the probe pulse gives rise to new features in the spectrum, such as new resonant profiles with zero or two angular nodes arise, the most prominent of which is the isotropic line at -3.3~eV below the N=2 threshold, corresponding to the well known $2p^2$ {$^1$S$^e$} resonance. Other prominent pump-probe features are the multiphoton sidebands of the {$^1$P$^o$} DESs, which interfere with the non-resonant one-photon ionization amplitude giving rise to group of fringes such as the one at -7~eV below the $N=2$ threshold.

Using a moderately intense IR probe pulse in association with an XUV pulse that brings the system to the region of the doubly excited states is an efficient way of populating doubly excited states with symmetries other than {$^1$P$^o$}.  This is because, in contrast to excitations from the ground state, radiative transitions between doubly excited states require to change the configuration of just a single electron. Indeed,
doubly excited states of {$^1$S$^e$} and {$^1$D$^e$} symmetry are clearly visible in the spectra in Fig.~\ref{fig:TelescopicSeries}. Furthermore, the dipole propensity rules observed for one-photon transitions from symmetrically excited states~\cite{Rost1990,Rost1991a}, and in particular from the ground state, do not apply to transitions between doubly-excited states. XUV-pump IR-probe spectroscopy, therefore, is a valid way of detecting DES with arbitrary symmetry, alternative e.g. to the measurement of non-dipole effects in one-photon ionization~\cite{Argenti2009,Krassig2012}.

As explained in Sec.~\ref{sec:theory}, the photoelectron distribution is computed by projecting the wave packet on a set of field-free multichannel scattering states. Since these states are eigenstates of the Hamiltonian, their population does not change with time, which means that the projection can be taken at any time between the end of the external pulses and the time at which the electrons in the channel and energy range of interest reach the absorption boundary of the quantization box. The time-invariance of the photoelectron spectrum with respect to the time of projection is a major feature of the present approach, since it does not require to wait for all the relevant component of the wavefunction to reach large distances from the reaction center. 
\begin{figure}[hbtp!]
\includegraphics[width=\linewidth]{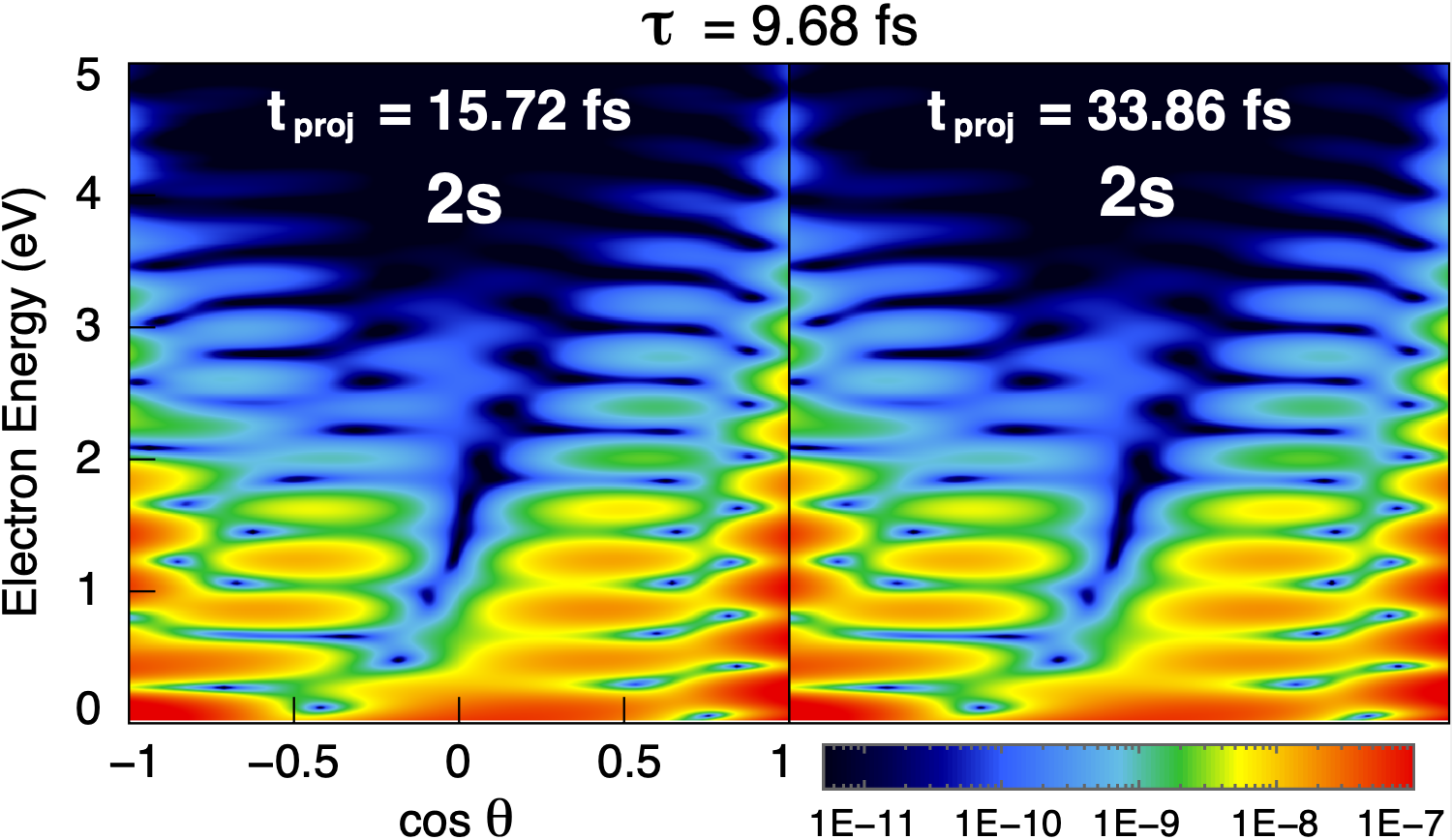}
\caption{\label{fig:d2P_dEdx_N2_Time_Invariance} {\bf Time invariance of the projection on scattering states:}Partial fully differential photoelectron distribution in the $2s$ channels, as a function of the photoelectron energy and ejection angle, for a sample pump-probe time delay $\tau=9.68$~fs (400~a.u.), obtained projecting the time-dependent wave packet at two different propagation times, $t_{\mathrm{proj}}=15.72$~fs (650~a.u.) and  $t_{\mathrm{proj}}=33.86$~fs (1400~a.u.).}
\end{figure}
Figure~\ref{fig:d2P_dEdx_N2_Time_Invariance} illustrates the time invariance of the final scattering-state population for the pump-probe simulation with XUV-pump IR-probe delay $\tau=9.68$~fs, carried out in velocity gauge. The figure shows the photoelectron distribution in the $2s$ channel, differential in both energy and photoemission angle, computed by projecting the time-dependent wavepacket on the scattering state at two different times: 1) immediately after the end of the IR pulse,  $t_{\mathrm{proj}}=15.72$~fs, and 2) at a much later time,  $t_{\mathrm{proj}}=33.86$~fs from the center of the XUV pulse. The two distributions are virtually identical, which demonstrates the effectiveness of this approach.

In time-dependent calculations involving multi-photon and possibly non-perturbative transitions, gauge agreement requires not only convergence with respect to the configuration space in any given symmetry, but also convergence of the time-dependent wave packet expansion with respect to the total angular momentum, as well as of the time integration itself. Under the influence of an intense long-wavelength external field, the velocity of a free electron undergoes large oscillations. In length gauge, where canonical momentum $\vec{p}_i$ and velocity coincide, therefore, the angular momentum $\vec{\ell}=\vec{r}\times\vec{p}$ also makes large excursions~\cite{Cormier1995}. In velocity gauge, instead, the canonical momentum for a free electron is a good quantum number. As a consequence, the expansion with respect to $L_{\mathrm{max}}$ usually has a more rapid convergence in velocity than in length gauge.

The lack of radiative coupling between free-electron states in velocity gauge is reflected also in the coupling between helium parent-ion states. The $2s$ and $2p$ He$^+$ hydrogenic states have a finite dipole matrix element $\langle 2s | r_1 | 2p\rangle$ and, within the electrostatic approximation, they are degenerate. In an external static field, these two properties give rise to a strong polarization of the ion (anomalous Stark shift). For the same reason, in length gauge, an intense external field causes the state of the ion to periodically oscillate between the $2s$ and $2p$ states. When integrating the TDSE numerically in the length gauge, therefore, this oscillation is an additional source of error that can give rise to unphysical radiative coupling between $2s\varepsilon_\ell$ and $2p\varepsilon_{\ell}$ channels. This circumstance is to be contrasted with the calculation in velocity gauge, where all degenerate states are rigorously radiatively decoupled,
\begin{equation}
\langle\Psi_i|\vec{P}|\Psi_j\rangle = i(E_i-E_j)\langle\Psi_i|\vec{R}|\Psi_j\rangle.
\end{equation}

In the present calculation, a maximum total angular momentum $L_{\mathrm{max}}=9$ is sufficient to reach convergence with respect to this parameter, in both the length and velocity calculation. 
\begin{figure}[hbtp!]
\includegraphics[width=\linewidth]{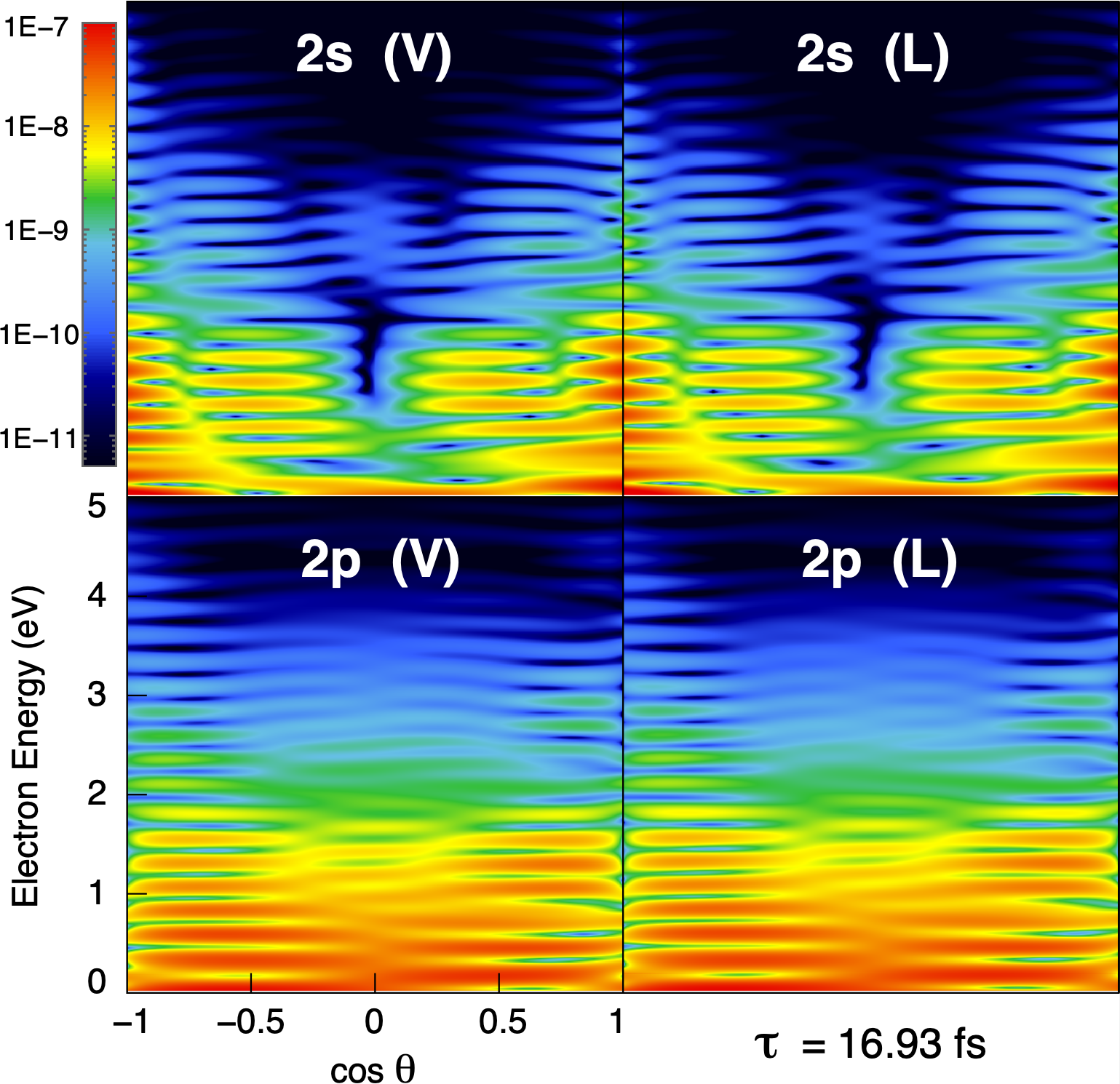}
\caption{\label{fig:d2P_dEdx_N2_vs_Gauge} Partial fully differential photoelectron distribution in the $2s$ and $2p$ channels, as a function of the photoelectron energy and ejection angle, for a sample pump-probe time delay $\tau=16.93$~fs (700~a.u.), computed in both velocity and length gauge. }
\end{figure}
\begin{figure*}[hbtp!]
\includegraphics[width=\linewidth]{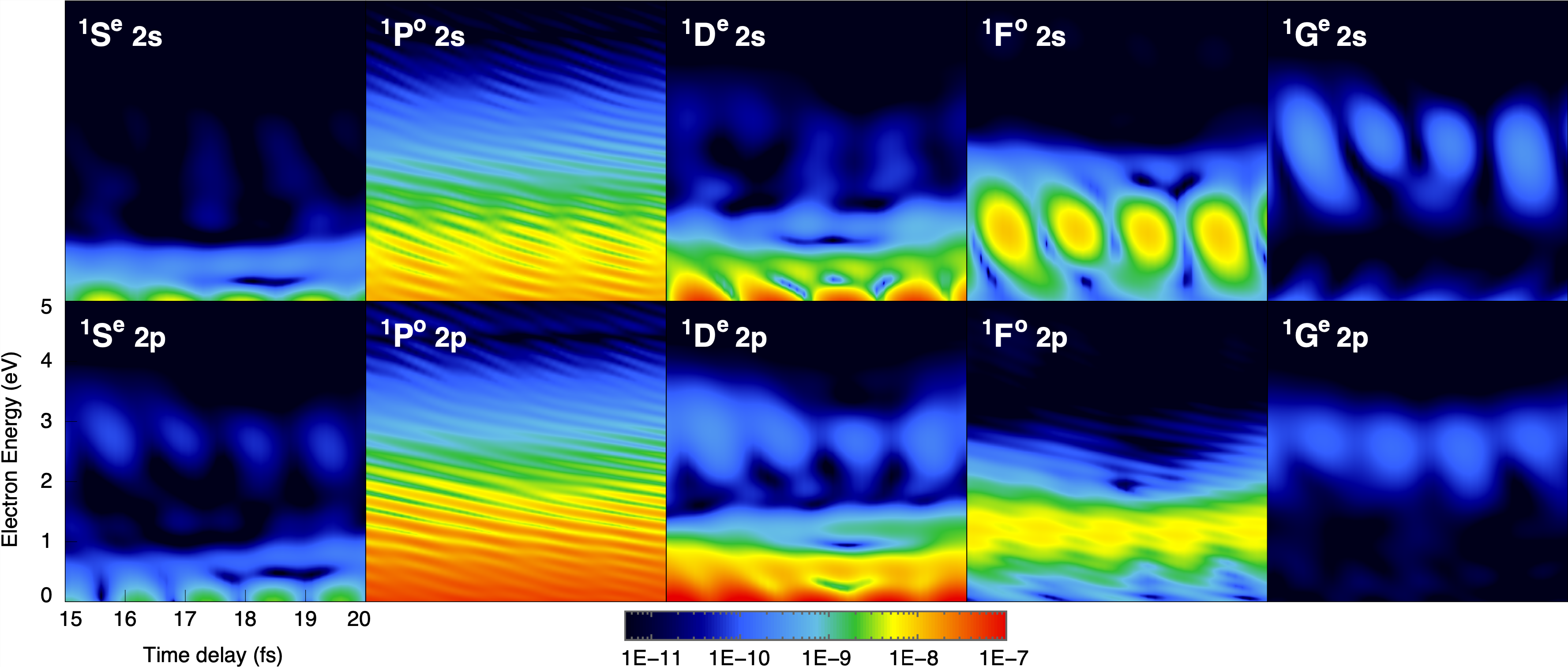}\vspace{0.5cm}
\includegraphics[width=\linewidth]{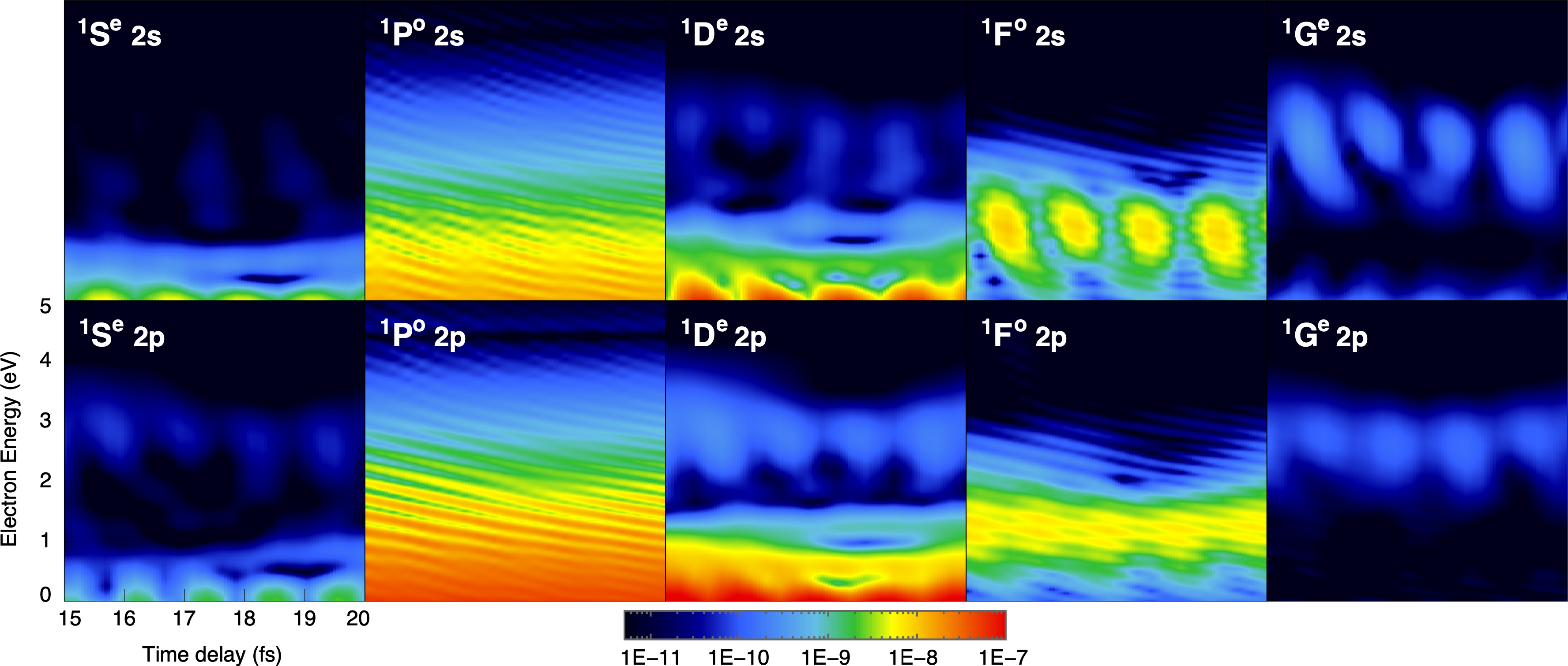}
\caption{\label{fig:dPaGdEvsT_VL_cor} Symmetry-resolved partial differential photoelectron distributions in the $2s$ and $2p$ channels as a function of the time delay and of the photoelectron energy in velocity gauge (top figure) and in length gauge (bottom figure). }
\end{figure*}
Figure~\ref{fig:d2P_dEdx_N2_vs_Gauge} compares the partial photoelectron distribution, fully differential in energy and emission angle, computed in the two gauges, in the case of the pump-probe time delay $\tau=16.93$~fs. The distributions are virtually identical in the two gauges, which is all the more impressive since the data are represented on a logarithmic scale. 

\begin{figure*}[hbtp!]
\includegraphics[width=\linewidth]{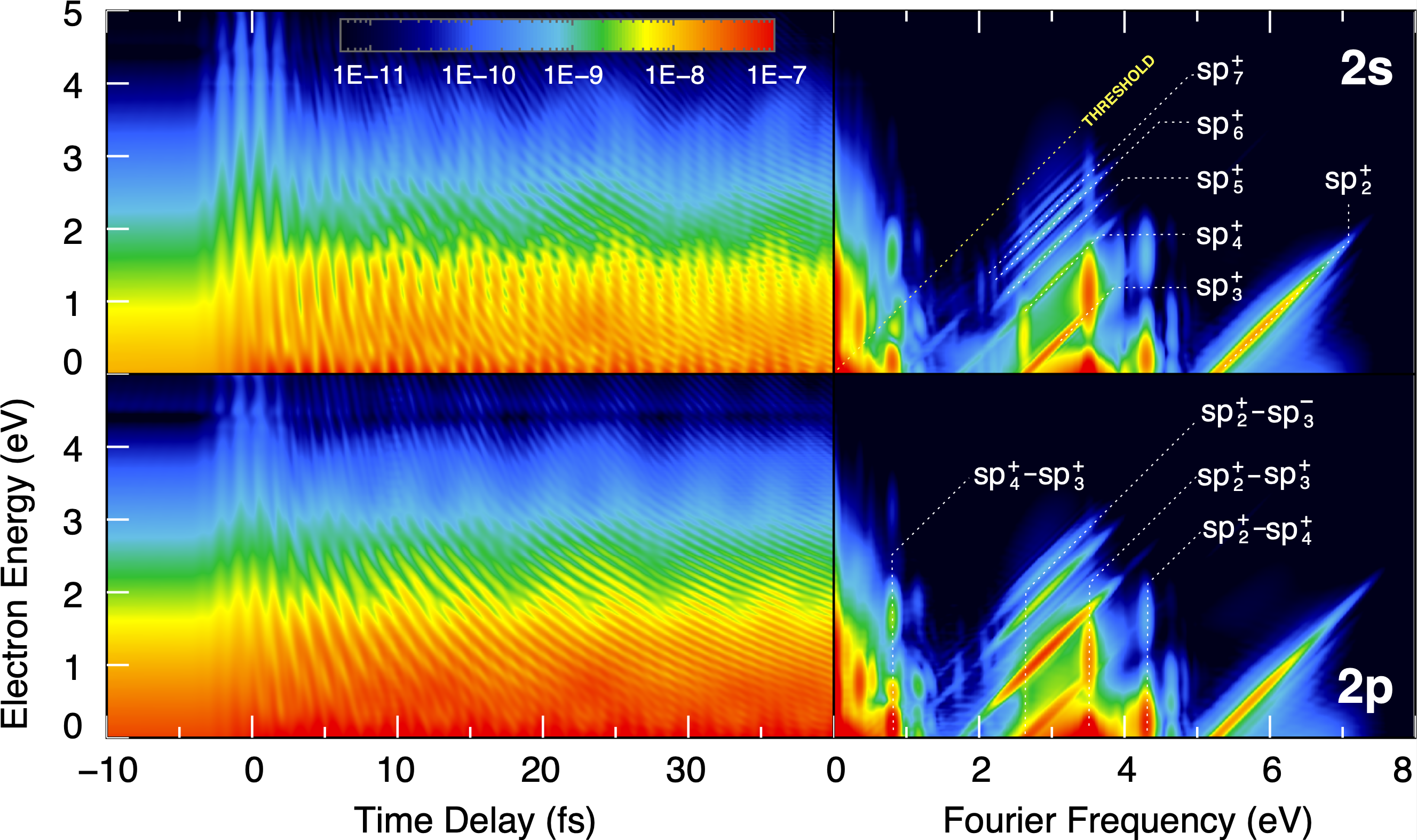}
\caption{\label{fig:dPadEvsT_cor} Partial differential photoelectron distributions in the $2s$ and $2p$ channels as a function of the time delay and of the photoelectron energy in velocity gauge. The Windowed Fourier Transform of the transient partial photoelectron spectrum with respect to the pump-probe delay reveals several slanted and vertical traces. The slanted traces, which are responsible for the characteristic hyperbolic fringes in the spectrum, are due to the beating between the direct shake-up one-photon amplitude, and the indirect ATI photoionization amplitude of the DESs. The vertical traces, instead, correspond to the beating between the ATI amplitudes from different intermediate DESs. Some of the most visible traces have been indicated in the picture.}
\end{figure*}
Figure~\ref{fig:dPaGdEvsT_VL_cor} shows the $2s$ and $2p$ photoelectron distributions, resolved by total symmetry and as a function of the pump-probe delay, computed in velocity and in length gauge. The calculations are clearly converged with respect to the total angular momentum. Indeed, the largest angular momentum with an appreciable population is $L=4$ ({$^1$G$^e$}). Furthermore, the {$^1$G$^e$} channels are overall an order of magnitude less populated than the {$^1$F$^o$} symmetry, and the $^1$G$^e$ partial spectra in length and velocity gauge are in excellent agreement.
The agreement between the two gauges is excellent for each symmetry and channel. The only exception is the $2s\varepsilon_{f}$ channel, which, in the length gauge, exhibits hyperbolic interference fringes that are a clear signature of contamination from the strongly populated $2s$ and $2p$ {$^1$P$^o$} channels. Whereas interference fringes are expected for the $2p\varepsilon_d$ {$^1$F$^o$} channel, due to its mixing with the $2p\varepsilon_d$ {$^1$P$^o$} channel induced by the dynamical polarization of the $2p$ parent ion, there is not any such transition mechanism that leads to the $2s\varepsilon_{f}$ states. The fringes predicted in length gauge for the $2s\varepsilon_{f}$, therefore, must be regarded as an indication of the superior numerical stability of the velocity gauge, for this system and for the propagator implemented here. The overall agreement between the two gauges is nonetheless excellent.

The use of a Ramsey-like spectroscopy with ultrashort XUV pulses to study coherent excitation of highly excited, and in particular of autoionizing states has been proposed already in 1998 by Cavalieri and Eramo~\cite{Cavalieri1998} and demonstrated experimentally on krypton autoionizing states shortly thereafter~\cite{Cavalieri2002}.
The interference fringes reported in Fig.~\ref{fig:dPadEvsT_cor} are the multi-channel equivalent of the phenomenon originally reported in 2010 by Anne L'Huillier and co-workers~\cite{Mauritsson2010}. These interferences have an holographic character, i.e., they result from the interplay between a pre-existing reference amplitude and a second signal which originate from the interaction of an electronic wave packet that evolves in time and an ultra-short probe pulse. From the interference with the reference, and by knowing some fundamental phases associated to the radiative transition from the different components of the wave packet to the final continuum, it is in principle possible to reconstruct the wavepacket and follow its evolution in time~\cite{Klunder2013}. The reconstruction following this approach, however, requires that the ionization process needed to ionize the dominant component of the wave packet is simple so that it can be described in terms of few analytical parameters. In the case of the doubly excited states of helium, however, the most populated state requires at least three IR photons to be brought above the N=2 threshold, through a resonant process, which introduces significant uncertainties in the reconstruction. 

\begin{figure}[hbtp!]
\includegraphics[width=\linewidth]{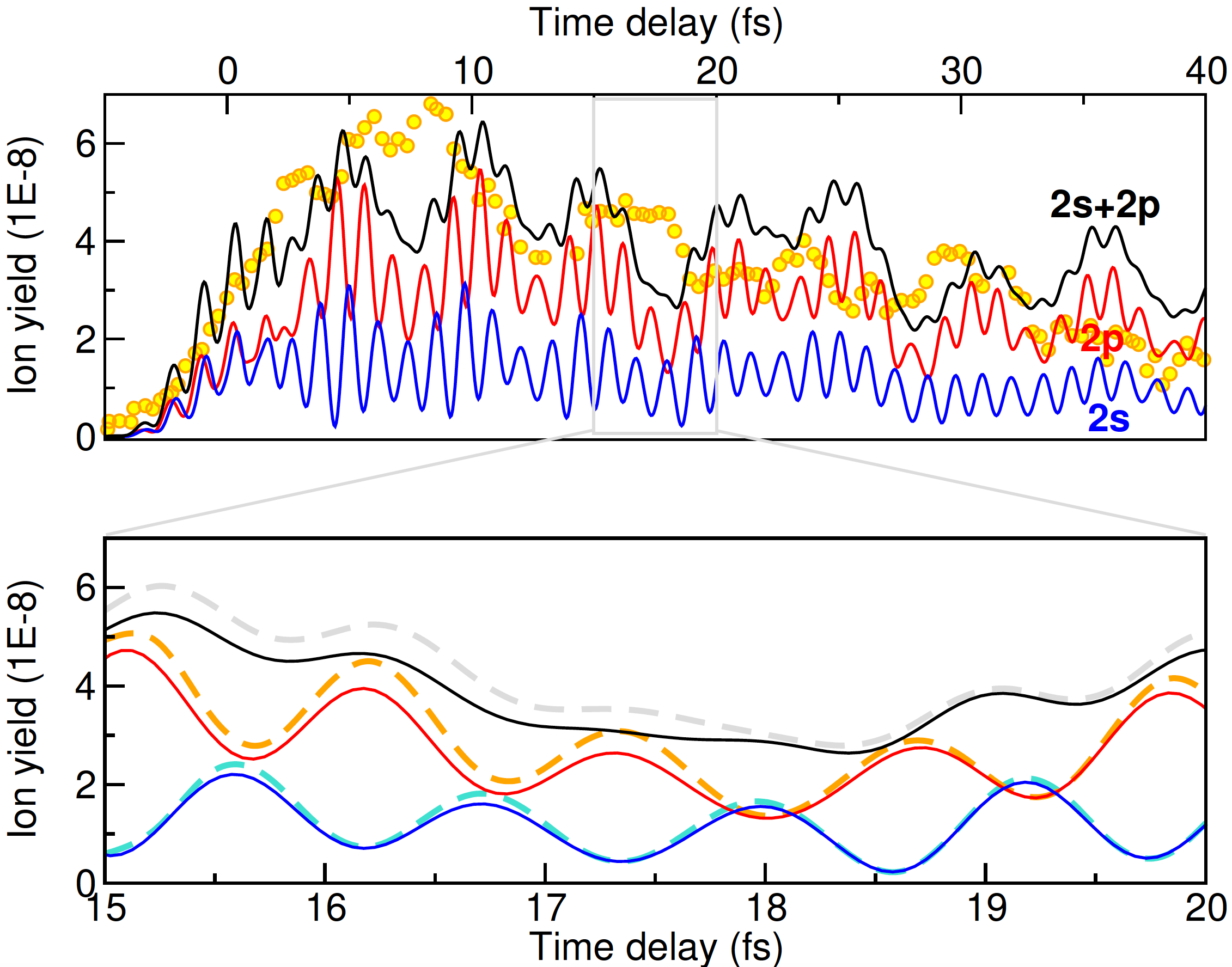}
\caption{\label{fig:PavsT_cor}Top panel: transient yields $\bar{P}_{a}(\tau)=P_{a}(\tau)-P_{a}(-\infty)$ of the He$^+$ parent ions, as a function of the pump-probe time delay. $\bar{P}_{2s}$ (lowest curve, blue online), $\bar{P}_{2p}$ (middle curve, red online), and $\bar{P}_{2\ell}=\bar{P}_{2s}+\bar{P}_{2p}$ (top curve, black online). The $2s$ and $2p$ yields oscillate almost in anti-phase. The yellow bullets represent the \emph{decrease} in the $1s\varepsilon_\ell$ photoelectron signal close to the $sp_2^+$ resonance in an experiment from 2010~\cite{Gilbertson2010}. The experimental data have been shifted in time, $\tau_{\mathrm{exp}}\mapsto \tau_{\mathrm{exp}}'=\tau_{\mathrm{exp}}+6$~fs, and the relative signal has been scaled to match the simulation. Bottom panel: comparison between the ionization yields computed in velocity (solid thin lines) and in length gauge (dashed thick lines), in a $5$~fs interval where the pump and probe do not overlap.}
\end{figure}

Figure~\ref{fig:PavsT_cor} shows the variation of the partial ionization yield in the $2s$ and $2p$ channels, computed using Eq.~\eqref{eq:partialintegralyield}. The $2\ell$ partial yields for a large negative time delay (only the XUV pump pulse has a role) are $P_{2s}(-\infty)=1.15[-8]$ and $P_{2p}(-\infty)=3.49[-8]$, irrespective of whether the calculation is performed in length or in velocity gauge. 
In this calculation, the $2s$ and $2p$ yields exhibit large oscillations at the frequency $E_{sp_3^+}-E_{sp_2^+}$ of the beating between the two dominant components of the metastable wave packet. These oscillations have comparable amplitudes for the two ions, and they are almost in antiphase. When considered together, the two modulations almost compensate and, as a consequence, the $N=2$ excitation probability becomes dominated by the beating between the $sp_3^+$ and the $sp_4^+$ state, the following term in the $sp^+$ series. When compared to the small-basis case shown in Fig.~\ref{fig:dPaGdEvsT_unc}, where the phase offsets between the $2s$ and $2p$ yields is just $\sim \pi/3$, these results demonstrate how dynamic electronic correlation and the polarizability of the parent ion have a dramatic impact on the ionization branching ratio. In particular, both electrons in the $sp_2^+$ and $sp_3^+$ DESs must be strongly affected by the dressing field. Furthermore, these results also suggest that the parent ion emerging from the ionization event is strongly polarized. 

Gilbertson~\emph{et al.}~\cite{Gilbertson2010} have realized a pump-probe experiment very similar to the one considered here, in which they monitored the signal of the fast electrons in the $1s$ channel, in the energy region close to the $sp_2^+$ state.
Whereas this observable is not equivalent to the $N=2$ yield, it is almost complementary to it. The signal in Gilbertson's experiment drops sharply in the transition from negative to positive time delays, to recover exponentially for larger time delays, with a characteristic time that coincides with the lifetime of the $sp_2^+$ resonance (see Fig.~2b in~\cite{Gilbertson2010}). The exponential recovery of the ionization signal was presented as the most prominent result of that work. A smooth exponential recovery of the signal was confirmed shortly thereafter in a theoretical paper based on a simplified few-level model~\cite{Chu2011}.  The $1s$ experimental resonant yield shown in Fig.~2b of~\cite{Gilbertson2010}, $P_{1s}(\tau_{\mathrm{exp}})$, which in the original paper is normalized to one, has been plotted in the top panel of Fig.~\ref{fig:PavsT_cor} as $A [ 1 - P_{1s}(\tau_{\mathrm{exp}}-6~\mathrm{fs}) ]$, alongside our theoretical prediction for the $N=2$ yield, where $A$ is a scaling constant. It is clear that, on top of the exponential profile, the experimental data exhibit also prominent broad fluctuations with frequency similar to that of the $sp_4^+\,-\,sp_3^+$ beating. Even in this rough comparison, the experimental fluctuations seem to mirror those observed in the complementary $N=2$ ionization-yield observable computed in the present work. It is possible, therefore, that the modulation in the experimental data in~\cite{Gilbertson2010} are the first experimental evidence of coherence in an autoionizing wave packet, rather than mere noise of instrumental origin, which would then predate by four years the reconstruction of the coherently excited $sp_{2/3}^+$ metastable wavepacket with optical methods~\cite{OttNature2014}. Additional investigations are needed to confirm this point. Our findings demonstrate the need of using accurate \emph{ab initio} representation of the system, whenever intense probe pulses are used. Indeed, simple analytical models~\cite{Zhao2005,Chu2011} or even minimal \emph{ab initio} close-coupling representations can miss important aspects of the ionization dynamics.

\section{Conclusions}\label{sec:conclusions}
We have shown how a time-dependent close-coupling code, in association with the projection on scattering states, is an ideal tool to reproduce the asymptotic photoionization distribution in the interaction of poly-electronic atoms with sequences of ultrashort laser pulses. The results reported here using a extensive correlation space show excellent agreement between length and velocity gauge in all the partial differential observables of interest. In particular, the original prediction of an off-phase oscillation of the $2s$ and $2p$ branching ratios in the shake-up XUV-pump IR-probe ionization of helium is not only confirmed, but shown to be even larger, once correlation is fully taken into account. 

\begin{acknowledgments}
This work was supported by the United States National Science Foundation under NSF Grant No. PHY-1607588, and NSF Grant No. 1912507.
We thank Stokes supercomputer, at the University of Central Florida, Mare Nostrum BSC, CCC-UAM (Centro de Computaci{\'o}n  Cient{\'i}fica, Universidad Aut{\'o}noma de Madrid) for allocation of computer time. L.A. acknowledges also funding from the European Research Council under the European Union's Seventh Framework Programme (FP7/2007-2013)/ERC grant agreement No~290853, the European COST Action CM0702, the ERA-Chemistry project No~PIM2010EEC-00751, the Marie Curie ITN CORINF, the MICINN project No~FIS2010-15127, ACI2008-0777 and CSD 2007-00010 (Spain). 
\end{acknowledgments}

\bibliography{biblio}

\end{document}